\documentclass[superscriptaddress,notitlepage,showpacs,prb,reprint,nofootinbib]{revtex4-1}

\usepackage{amsmath}
\usepackage{epsf}
\usepackage{graphicx}

\usepackage{dcolumn}
\usepackage{bm}

\usepackage{amssymb}
\usepackage{array}
\usepackage{varioref}
\usepackage{wrapfig}
\usepackage{etoolbox}
\usepackage{color}

\providecommand{\nn}{\nonumber}

\providecommand{\bv}[1]{\bm{\mathrm{#1}}}

\providecommand{\w}{\omega}
\providecommand{\W}{\Omega}
\providecommand{\q}{\bv{q}}
\providecommand{\Q}{\bv{Q}}
\providecommand{\p}{\bv{p}}

\renewcommand{\k}{\bv{k}}

\providecommand{\ve}{\varepsilon}
\providecommand{\ef}{\varepsilon_F}
\providecommand{\vf}{v_F}

\providecommand{\kf}{k_F}

\providecommand{\kf}{k_F}

\renewcommand{\q}{\bv{q}}

\providecommand{\gb}{\bar{g}}
\providecommand{\gpref}{\frac{\gb}{\tpp^3}}
\providecommand{\tp}{2\pi}
\providecommand{\tpp}{\left(2\pi\right)}

\providecommand{\Sg}{\Sigma}
\providecommand{\Sgt}{\tilde{\Sg}}

\usepackage{feynmp-auto}
\usepackage[caption=false]{subfig}
\usepackage{etoolbox}
\usepackage{mdframed}

\usepackage{graphicx}

\begin{document}
\title{Dynamical susceptibility near a long-wavelength critical point with a nonconserved order parameter}

\author{Avraham Klein}
\affiliation{School of Physics and Astronomy, University of Minnesota, Minneapolis. MN 55455}
\author{Samuel Lederer}
\affiliation{Department of Physics, Massachusetts Institute of Technology, Cambridge MA 02139}
\author{Debanjan Chowdhury}
\affiliation{Department of Physics, Massachusetts Institute of Technology, Cambridge MA 02139}
\author{Erez Berg}
\affiliation{Department of Physics, University of Chicago, Chicago IL 60637}
\author{Andrey Chubukov}
\affiliation{School of Physics and Astronomy, University of Minnesota, Minneapolis. MN 55455}

\begin{abstract}
  We study the dynamic response of a two-dimensional system of itinerant fermions in the vicinity of a uniform ($\Q=0$) Ising nematic quantum critical point of $d-$wave symmetry.
  The nematic order parameter is not a conserved quantity, and this permits a nonzero value of the fermionic polarization in the $d-$wave channel even for vanishing momentum and finite frequency: $\Pi(\q = 0,\W_m) \neq 0$. For weak coupling between the fermions and the nematic order parameter (i.e. the coupling is small compared to the Fermi energy), we perturbatively compute $\Pi (\q = 0,\W_m) \neq 0$ over a parametrically broad range of frequencies where the fermionic self-energy $\Sigma (\omega)$ is irrelevant, and use Eliashberg theory to compute  $\Pi (\q = 0,\W_m)$ in the non-Fermi liquid regime at smaller frequencies, where  $\Sigma (\omega) > \omega$.  We find that $\Pi(\q=0,\W)$ is a constant, plus a frequency dependent correction that goes as $|\W|$ at high frequencies, crossing over to $|\W|^{1/3}$ at lower frequencies. The  $|\W|^{1/3}$ scaling holds also in a non-Fermi liquid regime. The non-vanishing of $\Pi (q=0, \W)$ gives rise to additional structure in the imaginary part of the nematic susceptibility $\chi^{''} (q, \W)$ at $\W > \vf q$,   in marked contrast to the behavior of the susceptibility for a conserved order parameter. This additional structure may be detected in Raman scattering experiments in the $d-$wave geometry.
\end{abstract}

\maketitle
\section{Introduction}
\label{sec:introduction}

The behavior of strongly-correlated fermions in the vicinity of a quantum critical point (QCP) is one of the most fascinating problems in many-body physics. A complex interplay of dynamics, correlations, and geometry lead to a wide array of phenomena, such as superconductivity beyond the Bardeen-Cooper-Schrieffer paradigm, non Fermi-Liquid (NFL) behavior, competing and interwined order parameters, among other effects. Today, it is widely believed that many complex materials, most prominently the cuprate and iron-based high T$_c$ superconductors, are examples of such critical systems.

A traditional way to treat the physics near a
QCP is to study an effective low-energy  model of itinerant fermions coupled to near-critical order parameter fluctuations. Within this model, one can study how soft bosons affect fermionic properties, like the quasiparticle residue and lifetime. At the same time one can also study how gapless fermionic degrees of freedom affect the bosonic properties of a system, such as critical temperatures and scaling dimensions of order parameter fields.

The subject of this paper is the bosonic dynamics that appears as a result of the coupling to fermions. Specifically, we study a system of  fermions in two spatial dimensions coupled to fluctuations of a $d-$wave nematic order parameter $\phi$ near a critical point, at which $\phi$ orders.
Our  goal is to understand fermion-induced  dynamics of the $\phi$ field  near such a transition.  This dynamics is encoded in the $d-$wave fermionic polarization $\Pi (q, \W)$. In the bulk of the paper we study $\Pi (q,\W)$  as a function of Matsubara frequency $\W_m = 2\pi m T$. We also discuss the imaginary part of the nematic susceptibility in real frequencies towards the end of the manuscript.

At high temperatures, thermal fluctuations dominate, and the largest term in $\Pi(\q, \W_m)$ is the one with $\Omega_m=0$, so that the dynamical properties are frozen. As the temperature is lowered, quantum fluctuations become important and eventually, at $T=0$, $\W_m$ becomes a continuous variable. Then it is necessary to describe response functions in their full momentum-frequency space. We address the question of what is the magnitude and the frequency dependence  of $\Pi (q=0, \W_m) $ at low temperature, $T\to 0$.

The limit of $q=0$ and finite $\W_m$ has attracted far less attention than the opposite limit $\W_m \ll \vf q$ (see e.g. Refs. \onlinecite{Hertz1976,*Millis1992,*Millis1993,*Altshuler1994}). There are several reasons for this. First, most theories of quantum critical phenomena in metals predict a dynamical exponent $z>1$, so that the scaling regime is accessed for $\W_m\sim q^z\ll \vf q$. Second, if the order parameter is conjugate to a conserved quantity (e.g., it couples to total fermionic density or spin), the fermionic polarization $\Pi (q=0, \W_m)$ vanishes identically by the conservation law and, by continuity, must be small for $\W_m\gg \vf q$.

However, recent years have seen an increasing interest in anisotropic transitions, such as long wavelength nematic QCPs with a $d-$wave order parameter, which we study in this work.  This order parameter  couples to the $d-$wave component of fermionic density, for which  the polarization is not constrained by the conservation law, so nontrivial dynamics in the regime $\W_m \gg \vf q$ are indeed possible.  The regime $\W_m \gg \vf q$  can be probed in numerical simulations and is also accessible in Raman scattering experiments.
 Nonzero dynamic response at vanishing $q$ have also been detected in neutron scattering near ferromagnetic QCPs in several Ur compounds ~\cite{Huxley2003,Stock2011,Chubukov2014}, although we do not explicitly discuss this case here.

In this paper we compute $\Pi (q=0, \W_m)$ at a nematic QCP. We work at weak coupling and with a large number of fermionic flavors $N$. We present results appropriate to several parametrically broad regimes of frequency. There are two relevant frequency scales in the problem (expressions for which will appear in the next section), both much smaller than the Fermi energy $\ef$. The first, $\omega_1$, is the frequency  below which the Landau damping of the bosonic degrees of freedom by the fermions becomes important. The second scale $\omega_0\ll \omega_1$, is the one below which the self-energy  of the fermions becomes important and the system develops NFL behavior. Schematically, our results are
\begin{align}
\Pi(\q = 0,\W_m)\approx\begin{cases}\text{const.}+|\W_m|,\qquad &\omega_1\ll|\W_m|\ll \ef\nn\\
\text{const.}+|\W_m|^{1/3},\qquad &\omega_0\ll|\W_m|\ll \omega_1\nn\\
\text{const.}+|\W_m|^{1/3},\qquad &|\W_m|\ll \omega_0
\end{cases}
\end{align}
We emphasize that the frequency dependence of $\Pi(\q = 0,\W_m)$ does not change around $\W_m =\omega_0$, i.e., it is not modified when the system enters the NFL regime below $\omega_0$.

\begin{figure}
  \centering
  \includegraphics[width=\hsize,clip,trim=0 150 0 150]{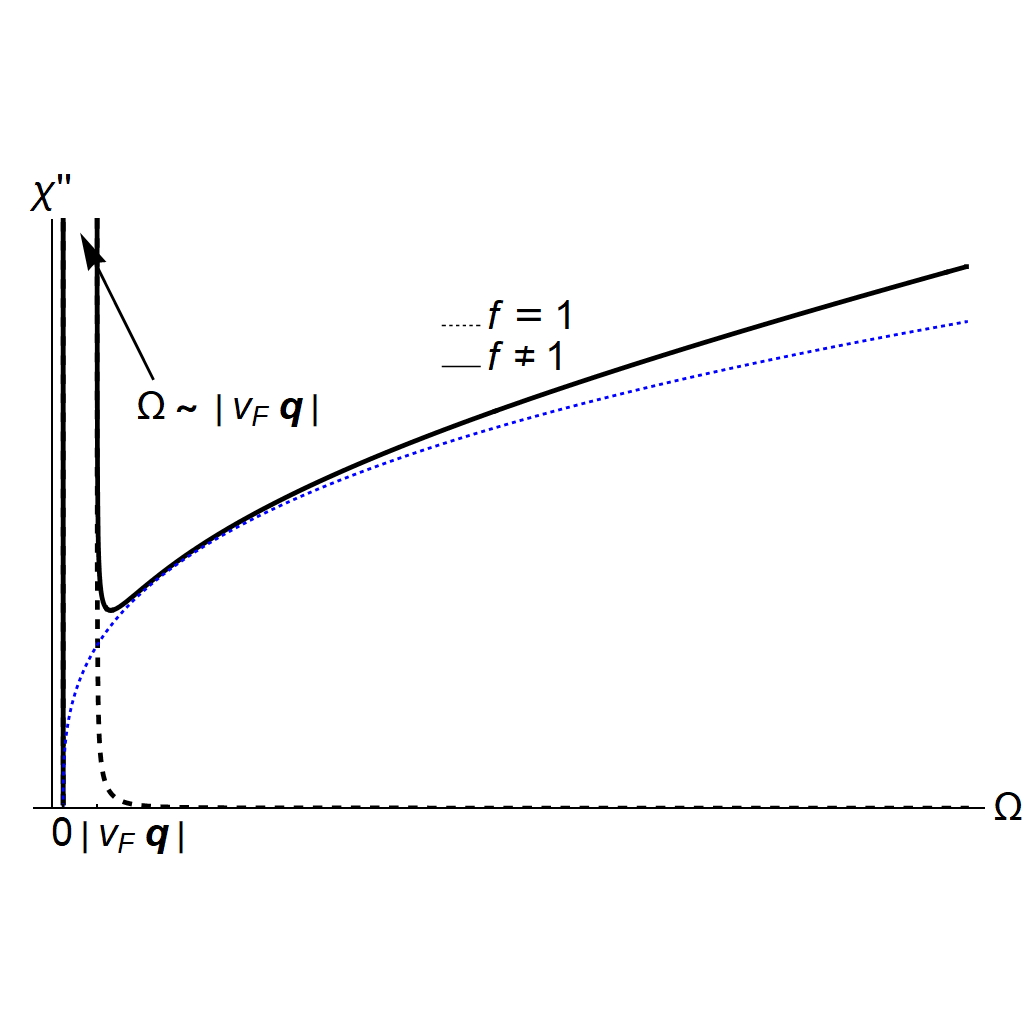}
  \caption{An illustration of the susceptibilities for a conserved vs. non-conserved order parameter. The figure is a sketch of the imaginary part of the susceptibility $\chi''(\q,\W)$ of collective excitations of a system of itinerant fermions near an isotropic QCP (dashed line), such as a ferromagnetic QCP, vs. a nematic QCP (solid line). In the low frequency regime $\W \ll \vf |\q|$ both susceptibilities are roughly identical, with a sharp peak at $\Omega \sim \vf |{\bf q}|$ due to Landau damping of the excitations. At higher frequencies $\W \gg \vf |\q|$, the isotropic response vanishes due to the conservation law (e.g. spin conservation). The nematic susceptibility flattens and then rises as $\W^{1/3}$ (dotted reference line), then switches to $\W$, before finally beginning to decay at $\W \sim \ef$ (not shown).}
  \label{fig:susc}
\end{figure}

In each of the three regimes, the frequency dependence is a small correction to the constant part. However, this frequency dependence
determines the imaginary part of the nematic susceptibility
in real frequencies, $\chi'' (q, \W)$, which scales as $|\W|^{1/3}$ for $|\W| \ll \omega_1$ and as $|\W|$ for $\ef\gg |\W| \gg \omega_1$ (see Fig. \ref{fig:susc}). This frequency dependence can be probed, for example, by Raman scattering\cite{Thorsmolle2016}.

The rest of this manuscript is organized as follows. In Sec. \ref{sec:theor-model} we introduce our model for a nematic QCP, give some motivation for the idea that $\Pi(\q=0,\W_m) \neq 0$ in this model, and derive the energy scales $\omega_0$ and $\omega_1$. In Secs. \ref{sec:incons-bare-pert}, \ref{sec:intermediate},
and \ref{sec:breakd-eliashb-theor} we present the calculations of $\Pi (q=0,\W_m)$, appropriate to the three regimes described above. We follow in Sec. \ref{sec:Raman} with
the analysis of nematic susceptibility and qualitative predictions for Raman scattering experiments, and present our conclusions in Sec. \ref{sec:discussion}. Technical details of the calculations are discussed in the Appendices.

\section{Model, general reasoning,  and energy scales}

\label{sec:theor-model}

In this section we introduce the model, present general reasoning why $\Pi (q=0, \W_m)$ should remain non-zero, and introduce relevant energy scales.

\subsection{The model}

We consider a two dimensional system with a scalar boson $\phi (q)$, which undergoes a continuous transition towards $d-$wave charge nematic order.
The bare susceptibility of the $\phi$ field is regular and can be approximated by
\begin{equation}
 D_0 (\q,\W_m) = \frac{\chi_0}{\xi^{-2}_0 +|\q|^2 + \W^2_m/c^2}
 \label{eq:chi0}
\end{equation}
where $\xi_0$ is the bare correlation length, which increases as the system approaches the QCP.
The dynamic $\W_m^2/c^2$ term is often neglected (though not always~\cite{Mahajan2013}), but we keep it.

We assume that there is a Yukawa coupling between $\phi(q)$ and $d-$wave fermionic density
\begin{equation}
  \label{eq:coupling-def}
  H_I = g \sum_{n=1}^{N}\sum_{\k,\q}f(\k)\phi(\q)\psi^{\dagger}_n(\k+\q/2)\psi_n(\k-\q/2).
\end{equation}
Here, $g$ is a coupling constant,  $n$ sums over the fermion flavors, and $f(\k)$ is a momentum dependent vertex with $d-$wave symmetry e.g., $f(\k) = \cos {k_x} - \cos{k_y}$.  Because our analysis is not too specific to $d-$wave symmetry of the nematic order, throughout the text we will keep $f(\k)$ as some function of momentum, without specifying its form.  We will use the $d-$wave form only at the end of calculations.

We assume that the fermions have a (not necessarily circular) Fermi surface (FS), dictated by
band structure. Below, we will only need $f(\k)$ for momenta near the FS, so we approximate $f(\k)$ by an angular function
\begin{equation}
  \label{eq:f-theta-def}
  f(\k) \approx
  f(\k = \kf \hat{\k}) \equiv f(\theta)
\end{equation}
where for a non-circular FS, $k_F$ by itself depends on $\theta$.

The effective fermion-boson model near a nematic QCP has been discussed before, but only in the regime where the characteristic frequencies $\W_m$ are small or, at most, comparable to $\vf q$. We will be interested in the properties of this model in the opposite limit, when $\W_m \gg \vf q$.

The full  susceptibility of the $\phi$ field  differs from $D_0$ due to the fermion-induced bosonic self-energy $\Pi(\q, \Omega_m)$:
 \begin{equation}
   D (\q,\W_m) = \frac{\chi_0}{\xi_0^{-2} +
     |\q|^2 + \W^2_m/c^2  +\gb \Pi (\q,\W_m)}
 \label{eq:chi}
 \end{equation}
 where $\gb = g^2 \chi_0$.   The quantity $\gb$  has dimensions of energy and can be viewed as the effective boson-fermion coupling constant. We work at weak coupling, meaning $\gb\ll\ef$. The  $\Pi  (\q,\W_m)$ in Eq. (\ref{eq:chi}) is the fully renormalized
particle-hole polarization bubble. The static part $\Pi(\q,0)$ contains a constant piece, which renormalizes $\xi_0$ into the true correlation length $\xi^{-2} = \xi^{-2}_0 +\gb \Pi (\q \to 0, \W_m =0)$,  and  a regular $\q^2$ term, which we just incorporate into the existing $\q^2$ term in (\ref{eq:chi}). The  dynamic part of $\Pi (\q,\W_m)$ contains Landau damping of the form $|\Omega|/(\vf |\q|)$ at $|\W_m| \ll \vf |\q|$, which  is a relevant perturbation near the QCP. Then, at $\Omega \ll \vf |\q|$,
\begin{equation}
  D (\q,\W_m) = \frac{\chi_0}{\xi^{-2} +|\q|^2 + \W^2_m/c^2  + \gamma \frac{|\W_m|}{\vf |\q|}}
 \label{eq:chi_1}
 \end{equation}
where $\gamma$ will be explicitly defined below (see Eq. (\ref{eq:pi-1-loop})). It has been demonstrated (see e.g. \cite{Lee1989,Altshuler1994,Rech2006}) that the characteristic $\Omega_m$ and $\q$, relevant for  the computation of the fermionic self-energy, do satisfy $|\Omega_m| \ll \vf |\q|$, i.e., in self-energy calculations one should use $D(\q,\W_m)$ given by (\ref{eq:chi_1}).

Our goal is to obtain the fermionic polarization and the nematic susceptibility in the opposite regime of vanishing $q$ and finite $\W_m$, at the low temperature limit $T \to 0$. We argue that $\Pi (\q=0,\W_m)$ is non-zero because there is no conservation law for $d-$wave fermionic polarization. We directly compute $\Pi (\q=0,\W_m)$  using a diagrammatic technique, starting from a particle-hole bubble of free fermions, and adding self-energy and vertex corrections to the bubble. We show that characteristic internal bosonic momenta $\Omega'$ and $\q'$ still obey $\Omega' \ll \vf |\q'|$, even when external $\q$ vanishes and external $\W_m$ stays finite.  This will allow us to use Eq. (\ref{eq:chi}) for propagators of bosons which dress particle-hole polarization bubble. By the same reasoning, we will use self-energy for fermions in the bubble, which is we obtain using the same  Eq. (\ref{eq:chi_1}).

\subsection{The polarization bubble, general reasoning}

For free fermions and at small momentum $q$, the polarization
$\Pi^{(0)} (q,\W_m)$ is  given by
\begin{align}
  \label{eq:polarization_1}
  \Pi^{(0)} (\q,\W_m) & =  \frac{k_F}{\pi \vf}  \int \frac{d\theta}{(2\pi)}
  f^2 (\theta) {\tilde \Pi}^{(0)} (\q, \W_m, \theta) \nonumber \\
  \tilde{\Pi}^{(0)} (\q, \W_m, \theta)  &=
                                          \frac{i\W_m - \vf q \cos{\theta}}{\vf q \cos{\theta} },
\end{align}
where $\vf$ is the Fermi velocity, which for a non-circular FS also depends on $\theta$. This form is non-analytic, i.e., the value of $\Pi^{(0)} (\q,\W_m)$  at $\W_m, \vf q  \to 0$ depends on the order in which the two variables go to zero. At $\W_m \ll \vf q$,  $\Pi^{(0)} (\q,\W_m) \sim ({\bar g} k_F/\vf) |\W_m|/(\vf q)$, up to a constant.  In real frequencies, this accounts for Landau damping. In the opposite limit, $\Pi^{(0)} (\q=0, \W_m)$ vanishes no matter what $f(\theta)$ is.  This vanishing can be understood by noticing that at $\q=0$ and small but finite $\W_m$, ${\tilde \Pi}^{(0)} (0, \W_m, \theta)$ coincides with the correlator of the total number of fermions along a particular direction in coordinate space, taken at different times. For free fermions, the number of fermions along any direction in space is separately conserved (because free particles do not scatter), hence the integrand for $\Pi^{(0)} (0,\W_m)$ vanishes even before integration over $\theta$.

This vanishing, however, does not hold once we include interactions. To see why this is so, consider a model of spinless fermions, define a quadrupolar density $n_f(\q) = \sum_{\k} \psi^\dagger(\k+\q/2)\psi(\k-\q/2)f(\k)$, and take a local interaction between these quadrupolar densities:
\begin{align}
  \label{eq:ff-model}
  H &= \sum_{\k} [\ve(\k)-\mu] \psi^\dagger(\k)\psi(\k) + H_1, \\
  H_1 &= g \sum_{\q} n_f(\q)n_f(-\q),
\end{align}
Let us compute the Heisenberg equation of motion for $n_{f}(0)= \sum_{\k} \psi^\dagger(\k)\psi(\k)f(\k)$. For free fermions, $g = 0$, and we trivially obtain,
\begin{align}
  \label{eq:nf-dot-free}
  d(\psi^\dagger(\p)\psi(\p)f(\p))/dt=0,
\end{align}
i.e. the number density of \emph{each} $\p-$state is separately conserved. Once we turn on the interaction term, separate $\p-$states will no longer be conserved. For a generic $f (\p)$ we find,
\begin{align}
  \label{eq:nf-non-cons}
  &i\dot{n}_{f}(\q=0) = [n_{f}(0), H_1] \nn\\
  &\propto\sum_{\k,\p,\q} \psi^\dagger(\k+\q/2)\psi^\dagger(\p-\q/2)\psi(\k-\q/2)\psi(\p+\q/2)\nn\\
  &\qquad \times f(\k)f(\p) \times \nn\\
  &[f(\k+\q/2)-f(\k-\q/2)-f(\p+\q/2)+f(\p-\q/2)]
\end{align}
Thus, generically, only for $f(\q)=1, f(\q)=\q$ is the R.H.S. equal to zero, as expected for
density and momentum conservation \cite{Kiselev2017}.
For any other form factor, we can expect some time-dependent behavior. Because the full $\Pi (\q=0, \W_m)$ is related to the correlator of $n_f (\q=0)$, the time dependence of $n_f(\q=0)$ will induce dynamics of $\Pi (\q=0, \W_m)$. These dynamics are precisely the topic of our work.

\subsection{Energy scales}

As discussed in the introduction, the model has
three parametrically broad regimes of frequency at weak coupling.
All three regimes can be identified right at a QCP where the dressed correlation length $\xi$ diverges.

The scale $\omega_1$, where Landau damping effects become important, can be deduced by comparing the Landau damping term in $\Pi^{(0)}$ with $D_0$ taken near the mass shell, i.e., at $|\q| \approx \W_m /c$.
 Using
\begin{align}
D_0^{-1}= &\frac{
 \q^2+\W_m^2/c^2}{\chi_0}\sim \frac{
 \W_m^2}{c^2\chi_0}\nn\\
 \Pi^{(0)}\sim&  N \frac{k_F}{\vf}\frac{|\W_m|}{\vf|\q|}\approx N\frac{ck_F}{\vf^2}
 \end{align}
and setting $ \chi_0 D_0^{-1}\sim {\bar g} \Pi^{(0)}$, we obtain
 \begin{align}
\omega_1\sim\sqrt{N \gb \frac{c^3 k_F}{\vf^2}}
   = \left(\frac{c}{\vf}\right)^{3/2}\sqrt{2 N\gb\ef}
\end{align}
where we defined $\ef = \vf k_F / 2$. For frequencies well above $\omega_1$ we can approximate $D (q, \W_m)$ by the bare susceptibility $D_0 (q, \W_m)$. For frequencies  below $\omega_1$, we must incorporate $\Pi^{(0)}$ into the boson propagator, i.e., replace
$D_0$ with
\begin{align}
D(\q, \W_m)=&\frac{\chi_0}{
             \q^2+\W^2_m/c^2
             +
             {\bar g} \Pi^{(0)}}\nn\\
\approx& \frac{\chi_0}{\q^2+\gamma
\frac{|\W_m|}{\vf|\q| }},\quad \gamma
  \sim  \frac{N \gb k_F}{\vf}
 \label{ch999}
\end{align}
For frequencies below $\omega_1$, it is appropriate to carry out the diagrammatic calculation of polarization using $D$ for internal boson lines. The internal fermion lines can be taken as free fermion propagators $G_0$ down to a still lower frequency $\omega_0$, at which the dressing of fermions by bosonic fluctuations can no longer be neglected. To estimate $\omega_0$, we compare the inverse of the bare fermion propagator to the one loop fermion self energy
\begin{align}
\Sigma_1(\k, \omega_m)
 =
 &g^2 \int \frac{d^2q d \Omega}{(2\pi)^3} G_0(\k+\q, \omega_m + \W_m)D(\q,\W_m)\nn\\
\sim& \left(\frac{\gb^2}{N\ef}\right)^{1/3}|\omega_m|^{2/3},
\end{align}
where we have included only the most singular part of the self energy. Since the bare inverse propagator goes as $\omega_m$, setting $G_0^{-1}\sim\Sigma_1$ gives
\begin{equation}
\omega_0\sim\frac{\gb^2}{N\ef}
\end{equation}

For frequencies below $\omega_0$, the dressing of fermion propagators must be accounted for.
 We will also show that in this regime vertex corrections play an important role in the calculation of $\Pi (\q=0, \W_m)$,  as will be discussed in Sec. \ref{sec:breakd-eliashb-theor}.

 For ${\bar g} \ll \ef$, the hierarchy of energy scales is $\omega_0 \ll \omega_1 \ll \ef$.  This condition sets three distinct {\it low-energy} regimes for $\Pi (\q=0, \W_m)$:  $\omega_1 <  |\W_m| < \ef$ (Regime $\rm{I}$), $\omega_0 <  |\W_m| < \omega_1$ (Regime $\rm{II}$), and $|\W_m| < \omega_0$ (Regime $\rm{III}$).  Below we present calculations for each energy regime in turn.

\begin{figure*}
  \centering
  \subfloat[\label{fig:2-loop-a}]{
    \includegraphics[width=0.33\textwidth,clip,trim=120 480 150 120]{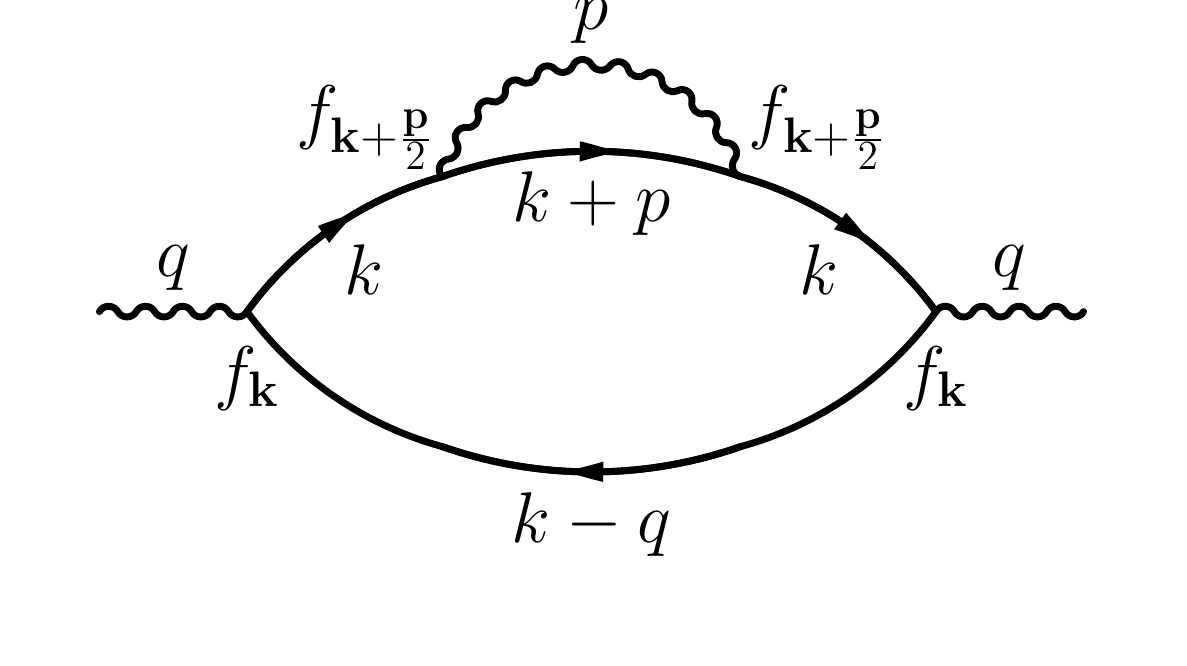}}
  \subfloat[\label{fig:2-loop-b}]{
    \includegraphics[width=0.33\textwidth,clip,trim=120 480 150 120]{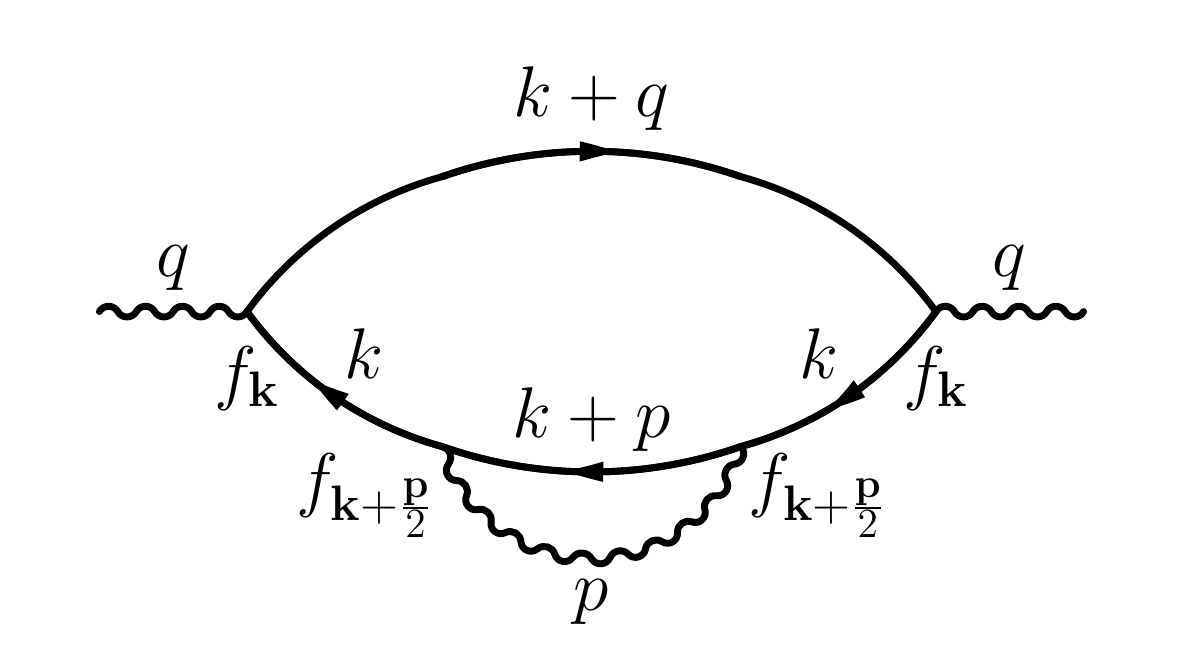}}
  \subfloat[\label{fig:2-loop-c}]{
    \includegraphics[width=0.33\textwidth,clip,trim=120 480 150 120]{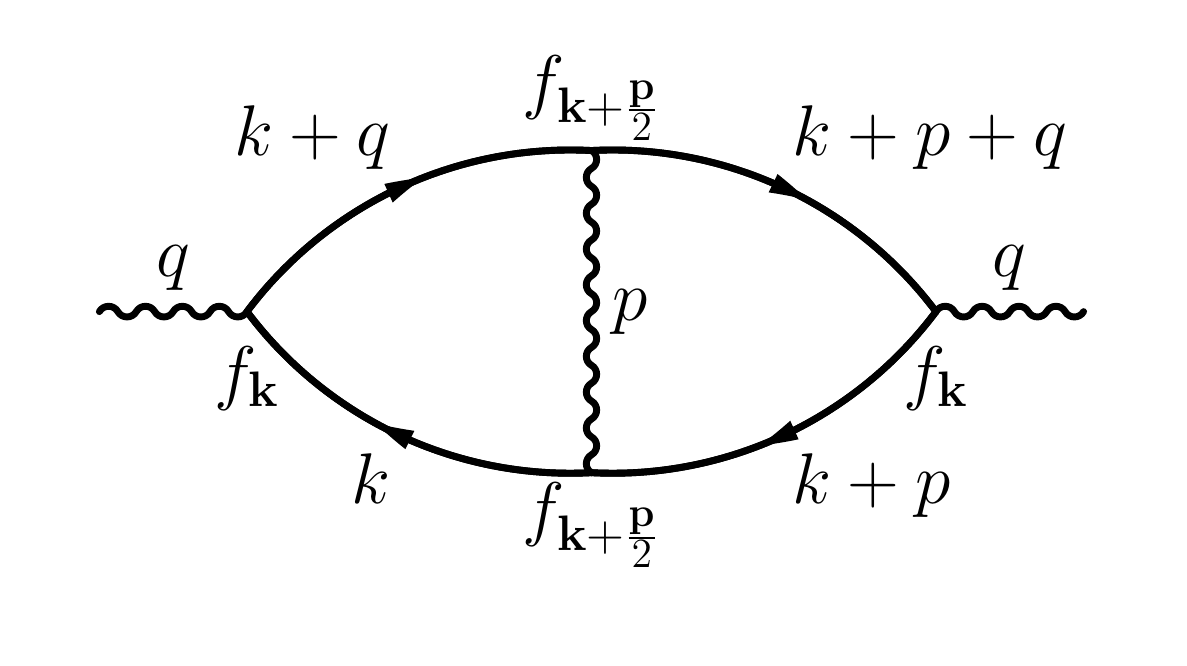}}
  \caption{The contributions to the polarization with one bosonic propagator inserted into a particle-hole bubble. The first two diagrams are self energy corrections and the last is the vertex correction. For a constant form-factor these three diagrams cancel exactly, as required by the Ward identity for number conservation.}
  \label{fig:2-loop}
\end{figure*}

\begin{figure*}
  \centering

  \begin{minipage}{0.4\hsize}
    \subfloat[\label{fig:AL-1}]{
      \includegraphics[width=\hsize,clip,trim=120 180 120 460]{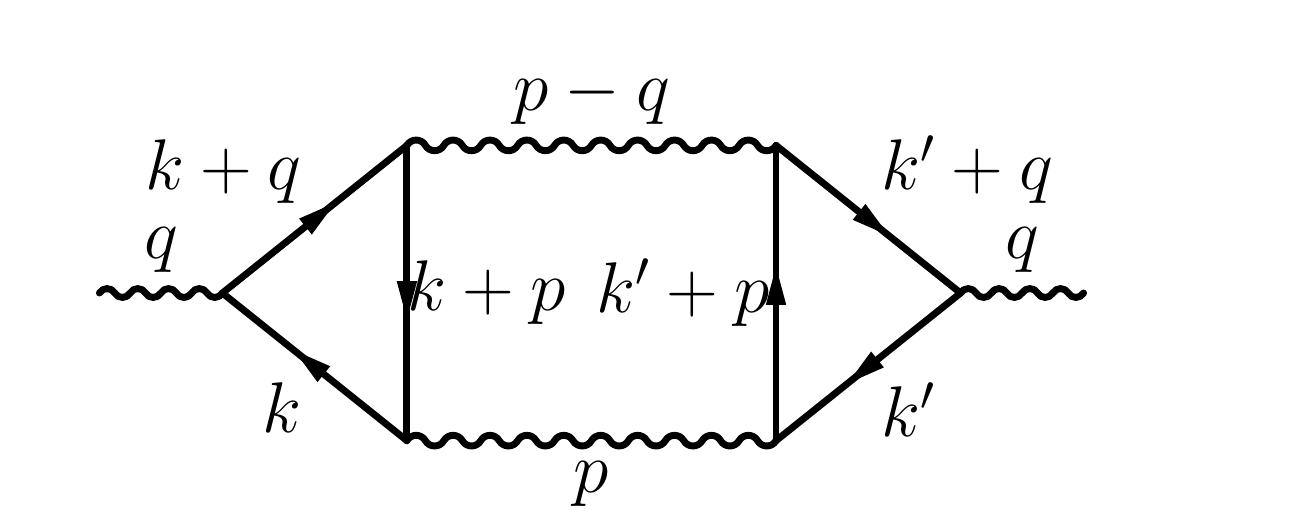}}
  \end{minipage}
  \begin{minipage}{0.4\hsize}
    \subfloat[\label{fig:AL-2}]{
      \includegraphics[width=\hsize,clip,trim=100 463 100 120]{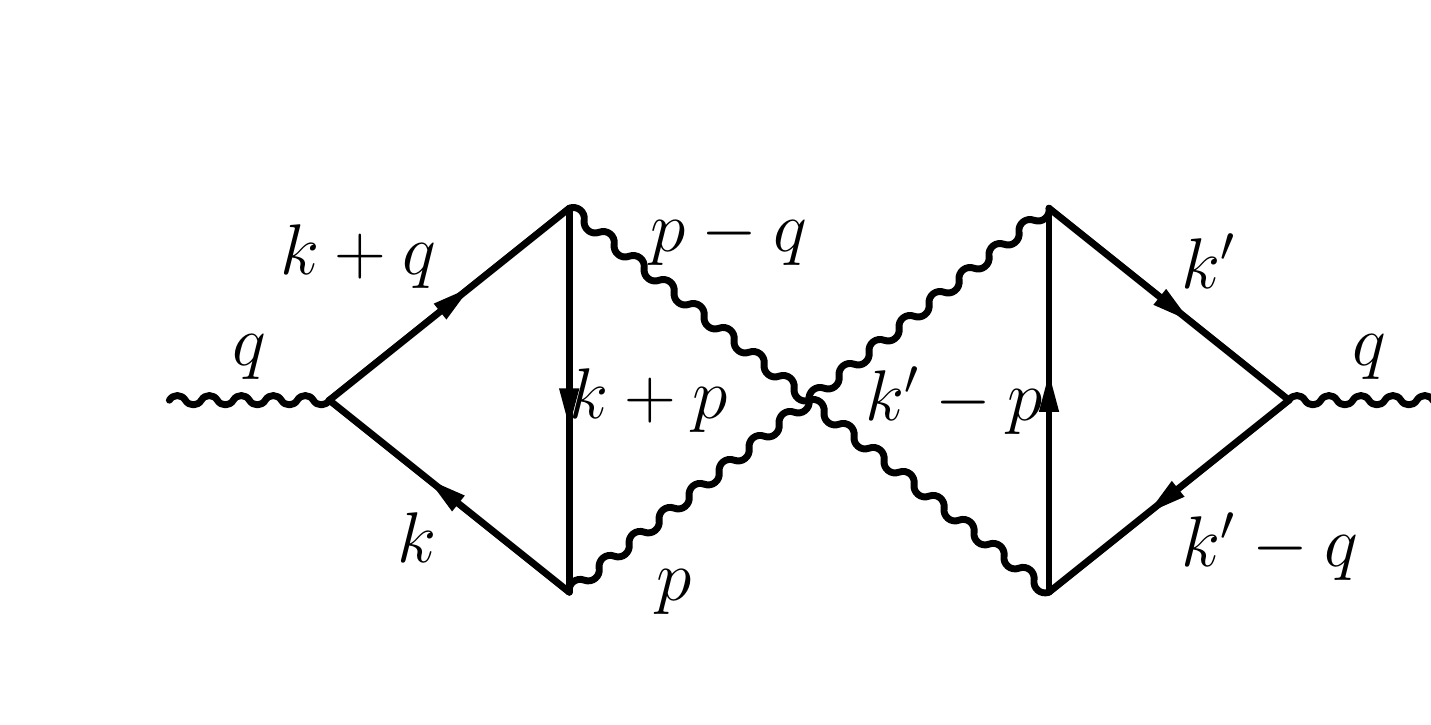}}
  \end{minipage}
  \caption{The leading contributions to the polarization with two bosonic propagators inserted into a particle-hole bubble (Aslamazov-Larkin diagrams). At frequencies $\W_m \ll \w_1$ these diagrams contribute to $\Pi$ at the same order as the diagrams of Fig. \ref{fig:2-loop}. For a constant form-factor, the two diagrams cancel exactly. For a non-constant form factor, calculating them mirrors the procedure for calculating the diagrams of Fig. \ref{fig:2-loop} (see Appendix \ref{sec:aslam-lark-corr} for details).}
  \label{fig:3-loop}
\end{figure*}

\section{Perturbative evaluation of the polarization $\Pi (\q \to 0, \W_m)$ in Regime $\rm{I}$}
\label{sec:incons-bare-pert}

To shorten formulas, in this and the following sections we will use three-vector notations for momentum and frequency: $q = (q_0, \q)$, $k = (k_0, \k)$.

We recall that for free fermions
\begin{equation}
  \label{eq:pi-1-loop=fk}
  \Pi^{(0)} (\q=0, q_0) = 0,
\end{equation}
even for $f(\k) \neq 1$. As noted earlier, this is because free fermions cannot exchange momentum, so the partial density of fermions for each direction of momentum is separately conserved. However, as we discussed in the previous section, there is no  reason to expect that $\Pi (\q=0, q_0) = 0$ will hold once we allow fermions to interact. We begin by evaluating the first nonzero contribution to $\Pi({\bf q}=0,q_0)$ within perturbation theory in the
 coupling $\gb$,
represented by the diagrams of Fig. \ref{fig:2-loop}. (See Appendix \ref{sec:deriv-results-sec-bare} for details.)

Let us consider the diagrams of Fig. \ref{fig:2-loop}. Each diagram contains four propagators of free fermions and one bosonic propagator. For a constant form-factor, these three diagrams cancel exactly, and the cancellation can be traced to the Ward identity for number conservation \cite{Baym1961}.
We show that for a non-conserved order parameter the three diagrams do  not cancel. In explicit form we have
\begin{equation}
  \label{eq:pi-2-bare}
  \Pi^{(1)}(q) = I_+ + I_- + I_v,
\end{equation}
where the diagrams with self energy insertions are
\begin{widetext}
\begin{align}
  \label{eq:I-pm-bare}
  I_\pm &= \frac{N\gb}{\chi_0\tpp^6}\int d^3k d^3p ~ G_0^2(k)G_0(k+p)G_0(k \pm q) D_0(p) f^2(\k \pm \q/2)f^2(\k+\p/2) \\
  \end{align}
while the diagram with a vertex correction is
\begin{align}
  \label{eq:I-v-bare}
    I_v &= \frac{N\gb}{\chi_0\tpp^6}\int d^3k d^3p ~ G_0(k)G_0(k + q)G_0(k+p)G_0(k+q+p)D_0(p) \times \nn \\
  &\qquad\qquad\qquad\qquad \times f(\k + \q/2)f(\k + \p + \q/2)f(\k+\p/2)f(\k + \q + \p/2)
\end{align}
Here, and henceforth, we replace the frequency sum by an integral, i.e. we assume $T \to 0$. We can recast these expressions into a more illuminating form by repeated application of the following identity of free fermion Green functions:
\begin{align}
  \label{eq:GG-kappa}
  G_0(k+p)G_0(k) = \mathcal{K}(k+p,k)[G_0(k)-G_0(k+p)]&\text{, where }\\
  \mathcal{K}(k+p,k) &= \left[
i p_0 - \ve(\k+\p) + \ve(\k)\right]^{-1} \nn\\
&\simeq \left[
i p_0- \vf \hat{k}\cdot \p\right]^{-1},
\label{eq:kappa-def}
\end{align}
Some straightforward algebra then yields:
\begin{align}
  \label{eq:Id-def}
  \Pi^{(1)} (\q=0, q_0) &= \frac{N\gb}{\chi_0\tpp^6}\int d^3k d^3p ~ G_0(k)G_0(k + q)G_0(k+p)G_0(k+q+p)D_0(p) f(\k)f^2(\k+\p/2) \times \nn\\
  &\qquad\qquad\qquad\qquad \times \left[f(\k+\p) - f(\k)\right]
\end{align}
\end{widetext}
We immediately see that for a constant $f$, $\Pi^{(1)}(\q=0, q_0)=0$, as it should, while for a momentum-dependent $f(\k)$, the two terms in the last bracket in (\ref{eq:Id-def}) do not cancel each other.

To estimate the value of the integral, we note that bosonic momentum ${\bf p}$ is naturally constrained by $k_F$, otherwise Eq. (\ref{ch999}) would not be valid.
  Approximating $f(\k+\p) - f(\k)$ by $|\p|^2$ and restricting integrations over $\k$ and $\p$ by $k_F$,  we obtain by power counting that
  $\Pi^{(1)} (\q=0, q_0)$ at
   $q_0 < \ef$ is a constant
    plus a subleading piece proportional to $|q_0|$:
\begin{equation}
  \label{eq:pi-perturbation}
  \gb\Pi^{(1)}(\q = 0, q_0) =  N\left(\frac{\bar g}{\ef}\right)^2 k^2_F \left(A
   + B \frac{|q_0|}{\ef}\right),
\end{equation}
where $A,B$ are dimensionless constants of order one, computed in the Appendix. The constant term is non-universal in the sense that it depends on the behavior of the system at bosonic momenta comparable to $k_F$.
   By contrast, the $|q_0|$ term is universal in the sense that it depends only on the form of the bosonic propagator at small momentum and small frequency.

\section{Evaluation of $\Pi (\q=0,q_0)$ in Regime $\rm{II}$}
\label{sec:intermediate}

To evaluate the polarization at frequencies comparable to or below $\omega_1$, the Landau damping of the boson must be explicitly incorporated. We now treat the parametrically broad regime  $\omega_0\ll q_0\ll \omega_1$, in which the damping of fermions can be neglected, but Landau damping of bosons plays a dominant role. In this regime, we can neglect the bare $q_0^2/c^2$ piece of the boson propagator in comparison with the Landau damping term. We again work perturbatively in the small parameter $\bar g/\ef$, but
use $D(q)$ given by (\ref{ch999}) instead of $D_0$ for the bosonic susceptibility.
In addition to the two-loop diagrams of Fig. \ref{fig:2-loop}, the Aslamazov-Larkin diagrams of Fig. \ref{fig:3-loop}
now yield contributions of the same order  (the extra overall factor of ${\bar g}$ in these diagrams is compensated by a $1/{\bar g}$ coming from the Landau damping).
However, they do not alter the qualitative result, and their treatment mirrors \cite{Chubukov2009,Maslov2010} that of the diagrams of Fig. \ref{fig:2-loop}. We defer their evaluation to Appendix \ref{sec:aslam-lark-corr}.

The power counting analysis of the integrals in Eq. \ref{eq:Id-def} is similar to that of the previous section, yielding
 the same constant part coming from momenta comparable to $k_F$.
 The frequency dependence, however, is altered by the new kinematics introduced by Landau damping, changing the exponent to $1/3$ instead of $1$. Explicitly:
\begin{equation}
  \label{eq:pi-bare-wrong}
  \gb\Pi^{(1)}(\q = 0, q_0) =  N\left(\frac{\bar g}{\ef}\right)^2 k^2_F \left(A  + C \left(\frac{ {N\bar g} |q_0|}{\epsilon^2_F}\right)^{1/3}\right),
\end{equation}
where $C$ is a  dimensionless constant of order one, computed in Appendix \ref{sec:deriv-results-sec-bare}. Similar to the result at $q_0\gg\omega_1$, the
frequency dependent piece is universal in the sense that it depends only on the form of the propagator at small frequency and momentum.

One can check that terms with larger number of bosonic propagators  are progressively small in ${\bar g}/\ef$ and hence irrelevant. As a result, the full $\Pi (\q=0, q_0)$  is well approximated by Eq. (\ref{eq:pi-bare-wrong}).

The scaling forms in the Regimes I and II, Eqs. (\ref{eq:pi-perturbation}) and (\ref{eq:pi-bare-wrong}), can be viewed as the limiting cases of a single scaling function of $q_0/\omega_1$. We present this function in Eq. \eqref{eq:supp-mu1-explicit-2} in Appendix \ref{sec:deriv-results-sec-bare}.

\section{Evaluation of the polarization in Regime $\rm{III}$: Eliashberg theory}

\label{sec:breakd-eliashb-theor}

We now move to frequencies $q_0 \lesssim \w_0$. Here we must account for both Landau damping and the large fermionic self-energy. Seemingly, we should proceed in this case the same way that we did in the previous section, by incorporating the self energy $\sim\omega^{2/3}$ into the fermionic propagator, $G^{-1}(k) = i \left(k_0+ \Sigma_1 (k)\right) -\ve(\k)$.
Such an approach brings up the issue of potential double counting in diagrams \ref{fig:2-loop-a}+\ref{fig:2-loop-b} in Fig. \ref{fig:2-loop}, but let us ignore this for a moment.

   The calculation of $\Pi^{(1)} (\q=0, q_0)$ with the full $G(k)$ proceeds in the same way as for free fermions, however now $\mathcal{K}(k+p,k)$ in Eq. (\ref{eq:GG-kappa}) takes the form
  \begin{align}
  \label{eq:kappa-def_1}
  \mathcal{K}(k+p,k) &= \left[ i\Sgt(k+p) - i\Sgt(k) - \ve(\k+\p) + \ve(\k)\right]^{-1} \nn\\
 &\simeq \left[  i\Sgt(k_0+p_0) - i\Sgt(k_0)- \vf \hat{k}\cdot \p\right]^{-1}
\end{align}
where
\begin{equation}
  \label{eq:sgt-def}
  \Sgt(k) = k_0 + \Sg(k).
\end{equation}
The  expression for $\Pi^{(1)} (\q=0, q_0)$ becomes
\begin{widetext}
  \begin{align}
    \label{eq:Id-def_a}
    \Pi^{(1)} (\q=0, q_0) &= \frac{N\gb}{\chi_0\tpp^6}\int d^3k d^3p ~ G(k)G(k + q)G(k+p)G(k+q+p)D(p) f(\k)f^2(\k+\p/2) \times \nn\\
                      &\qquad\qquad\qquad\qquad \times \left[f(\k+\p) - \mathcal{K}(k+q,k)\mathcal{K}^{-1}(k+q+p,k+p)  f(\k)\right]
  \end{align}
\end{widetext}
Analyzing this expression at  $q_0 < \omega_0$, we see that the self-energy gives rise to two effects. First, the term, which was a constant without self-energy, now becomes of order $q^{2/3}_0$. It still comes from bosonic momentum $|\p|$ of order $k_F$, but now the integral over the two fermionic dispersions and the two frequencies yields $(q_0/{\tilde \Sigma} (q_0))^2 \sim q^{2/3}_0$.  Second, the low-energy contribution remains of order $q^{1/3}_0$ because the extra $q^{2/3}_0$ from fermions is compensated by an additional $1/|\p|^2 \sim 1/q^{2/3}_0$, since the term $f(\k+\p) - \mathcal{K}\mathcal{K}^{-1} f(\k)$ in (\ref{eq:Id-def_a}) is now of order of one rather than of order $|\p|^2$.

We will see below that the  frequency dependence of $\Pi (\q=0, q_0)$ remains of order $q^{1/3}_0$ both above and below $\omega_0$. However, the statement that the constant term gets replaced by $q^{2/3}_0$ will not survive once we include vertex corrections. To see  that vertex corrections must be included along with dressing of fermionic lines by the self-energy we note that $\Pi^{(1)} (\q=0, q_0)$ in Eq. (\ref{eq:Id-def_a}) is non-zero even when the form-factor $f(\k)$ is a constant. This is obviously {\it incorrect} because an isotropic charge order parameter (the total density) is a conserved quantity.  A survey of the $s-$wave problem \cite{Chubukov2005} shows that at a QCP, vertex corrections are of order one at any order of perturbation theory, if one uses full propagators for fermions. The extension to large $N$ eliminates crossed vertex correction diagrams, but ladder vertex corrections still remain of order one\footnote{The extension to large $N$ also does not eliminate other planar non-ladder diagrams \cite{Lee2009,Metlitski2010a,Holder2015}, that are of leading order in $1/N$. We comment on this later in this section}.  To verify that the full $\Pi (\q=0, q_0)$ vanishes for $f=1$, one has to sum up an infinite ladder series of vertex corrections, so that an account of vertex corrections is crucial to yield sensible physics in the regime $q_0\ll\omega_0$. Naturally, similar corrections must also be accounted for in our case of non-constant $f(\k)$.

\begin{figure}
  \begin{align}
    \parbox{0.2\columnwidth}{
    \includegraphics[width=0.2\columnwidth,clip,trim=160 500 270 120]{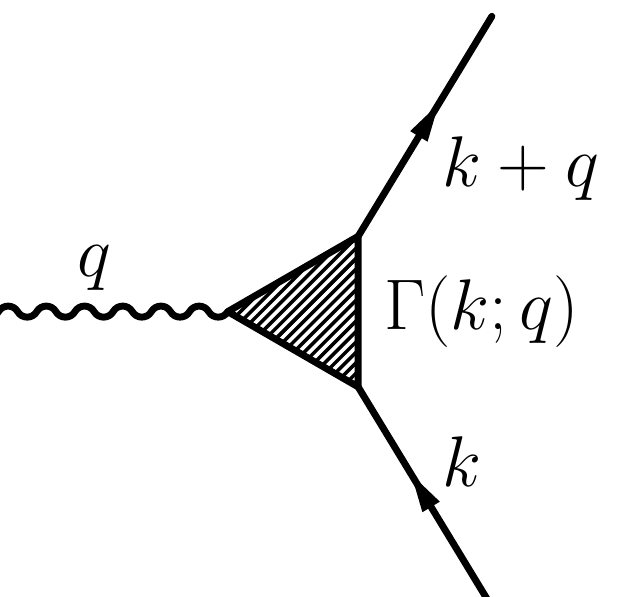}} &
                                                                                        \begin{array}{c} = \end{array}
  & \parbox{0.2\columnwidth}{\includegraphics[width=0.2\columnwidth,clip,trim=160 500 270 120]{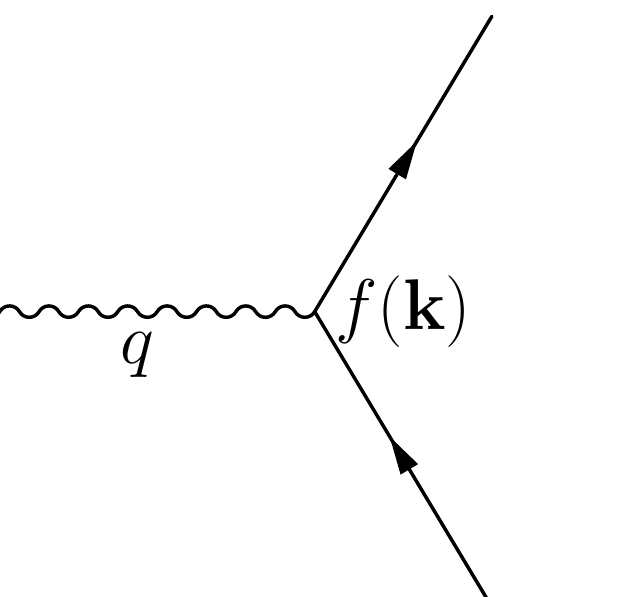}} & + &                                                                                       \parbox{0.2\columnwidth}{\includegraphics[width=0.2\columnwidth,clip,trim=160 500 270 120]{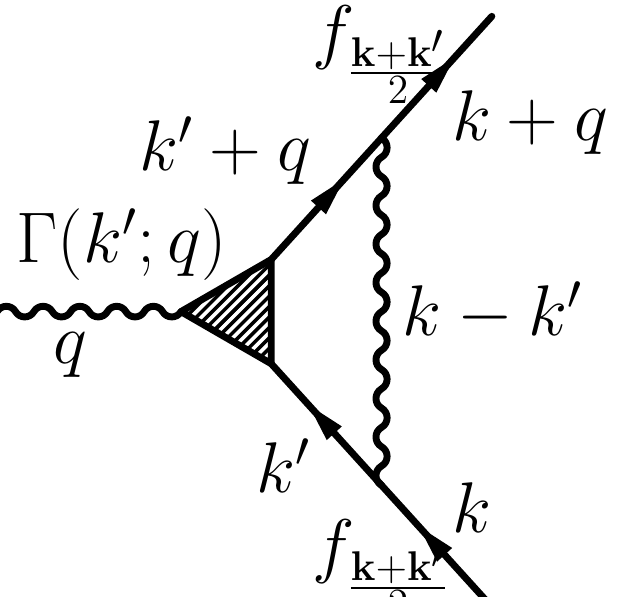}}
  \end{align}
  \caption{The coupling vertex. The leftmost panel depicts the fully renormalized vertex $\Gamma(k;p)$, where $k,p$ are 2+1 vectors. The two righthand panels are respectively the bare vertex in the fermion-boson model we define in sec. \ref{sec:theor-model}, and the vertex correction. The Green's functions and susceptibilities are full ones. In this work we adopt a ladder approximation for the vertex: We neglect crossing diagrams, and include self-consistent  self energy corrections in internal propagators. See Sec. \ref{sec:breakd-eliashb-theor}.}
  \label{fig:coupling-vertex}
\end{figure}

To analyze the vertex corrections,  we adopt a {\emph conserving} approximation \cite{Baym1961},
 meaning a choice of diagrams such that $\Pi(\q=0,q_0)$ vanishes for $f(\k=1)$. This approximation entails keeping the ladder series of vertex corrections pictured in Fig. \ref{fig:coupling-vertex}, but neglecting vertex corrections involving crossings. The kinematics of Landau damping will be of central importance to the calculations. The Landau damping term in the boson propagator means that the effective ``velocity'' of a collective boson is parametrically smaller than the Fermi velocity $\vf$. Thus, in any diagram which involves a fermion and a boson, depending on the same running momentum, one can factorize the momentum integration. One integrates over the momentum component transverse to the FS in a fermionic propagator, and over the momentum component along the FS in the bosonic propagator, neglecting there the momentum component along the FS. This is essentially the same physics that is incorporated in
Eliashberg theories  of quantum critical metals.\cite{Altshuler1994,Chubukov2005,Abanov2003,Rech2006}

We emphasize that the  conserving approximation is {\it not} a controlled  approximation in the usual sense of the word.
Although leading order corrections to ladder series of vertex renormalizations are small in $1/N$,
large $N$ does not in fact fully control the theory because some higher-order non-ladder vertex correction diagrams
are not suppressed by $1/N$
~\cite{Lee2009,Metlitski2010a}. Furthermore, the computation of certain four-loop diagrams for bosonic susceptibility~\cite{Holder2015} has cast doubt on the validity of $z=3$ scaling for the bosonic propagator.
Modifications of the problem\cite{Mross2010,Dalidovich2013,Fitzpatrick2014} to achieve mathematical control have been performed,
 as well as extensive Monte Carlo simulations \cite{Schattner2016,Lederer2017}, but no clear consensus has emerged\cite{Punk2016,Klein2017}.

\begin{figure}
  \includegraphics[width=\columnwidth,clip,trim=120 500 150 120]{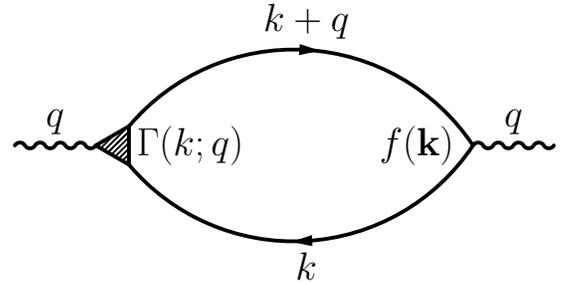}
  \caption{The fully dressed polarization bubble. For a boson coupled to a conserved quantity, the fully dressed polarization at $p = (p_0, \p=0)$ must be exactly zero due to the Ward identity. The bubble is dressed with the vertex depicted in Fig. \ref{fig:coupling-vertex}.}
  \label{fig:pol-bubble}
\end{figure}

With this caveat, we proceed with the conserving approximation. The perturbative series for the fully renormalized polarization bubble can be cast into the diagram shown in  Fig. \ref{fig:pol-bubble}, which expresses $\Pi (q)$ in terms of two dressed Green's functions and one  dressed vertex.  Each diagram in the perturbation series is counted only once, i.e., there is no double counting.  In explicit form we have
\begin{equation}
\label{eq:pi-def_1}
  \Pi(q) = N\int \frac{d^{3}k}{\tpp^3}\Gamma (k;q) G(k)G(k+q)f(\k+\q/2).
\end{equation}
The dressed fermion-boson vertex $\Gamma (k, q)$ is normalized such that for free fermions it reduces to $f(\k+\q/2)$.
The ladder diagrams for the vertex  $\Gamma(k;q)$ are shown in Fig. \ref{fig:coupling-vertex}.
We have verified that internal momenta and frequencies, which  mostly contribute to these diagrams at $q = (0, q_0)$, are the same as in Eq. (\ref{eq:char-mom-freq}). Accordingly, we will be using Eliashberg forms of bosonic and fermionic propagators: Landau-overdamped $D (q)$ from Eq. (\ref{ch999}) and dressed $G(k)$  with the self-energy given by Eq. (\ref{eq:se=1-loop}). We first demonstrate that $\Pi (q=0, q_0)$ indeed vanishes for a constant form factor due to particular cancellations between self-energy and vertex corrections, as specified by a Ward identity. Then we show that such a cancellation no longer holds for a non constant form factor, and,  as a result, find a nonzero $\Pi (q=0, q_0)$. Finally, we derive the same nonzero result in an alternative way, by analyzing the contribution given by each rung of a ladder diagram.

\subsection{Eliashberg theory}

Before delving into the full calculation involving vertex corrections, we present some
explicit results from the
Eliashberg theory for $q_0 \ll \vf |\q|$ (see Appendix \ref{sec:derivation-eqs-pi-sig} for details).  In this theory, the fermionic self-energy $\Sigma (k)$ depends on $k_0$ and on the position on the Fermi surface, but not on the momentum component transverse to the Fermi surface.  The theory is based on a set of self-consistent equations for the
polarization bubble (the bosonic self-energy):
\begin{equation}
  \label{eq:pi-def}
  \Pi(q) = N \int \frac{d^{3}k}{\tpp^3} G(k-q/2)G(k+q/2)f^2(\k),
\end{equation}
and the fermionic self-energy
\begin{equation}
  \label{eq:sigma-def}
  i\Sigma(k) = g^2 \int \frac{d^3p}{\tpp^3} G(k+p)D(p)f^2(\k+\frac{\p}{2}).
\end{equation}
In these equations,  $G(k \pm q/2)$  is the fermionic Green's function with self-energy included:
\begin{align}
  \label{eq:fermion-green-def}
  G(k) = \frac{1}{ik_0 + i\Sg(k) - \ve(\k) + \mu},
\end{align}
and  $D(q)$ is the bosonic susceptibility with the bosonic self-energy included
\begin{equation}
  D(q) = \frac{\chi_0}{\xi_0^{-2} + \q^2
  + q_0^2/c^2
  +\gb\Pi(q)}.
\end{equation}
Evaluating the momentum
integrals in Eq. \eqref{eq:pi-def}, we obtain that $\Pi (q)$ does not depend on the self-energy and has the same form as for free fermions~
(Ref. ~\cite{Oganesyan2001}):
\begin{equation}
  \label{eq:pi-1-loop}
  \gb\Pi(q) = \gamma \left(f^2({\hat {\bf q}}')\frac{|q_0|}{\vf |\q|} - \langle f^2 \rangle \right),\qquad \gamma = \frac{N\gb \kf}{\tp\vf},
\end{equation}
where
\begin{equation}
  \label{eq:angular-avg-def_1}
  \langle f^2(\theta) \rangle = \int \frac{d\theta}{\tp}f^2(\k =k_F {\hat {\bf k}}).
\end{equation}
and ${\hat {\bf k}}$ depends on the angle $\theta$ along the FS. Also, ${\hat z}$ is a unit vector in the direction perpendicular to the 2D plane, and ${\hat {\bf q}}' = {\hat z} \times {\hat {\bf q}}$, i.e., ${\hat q}'$ is orthogonal to ${\hat q}$. The dependence on ${\hat {\bf q}}'$ emerges because the momenta ${\bf k}$ in Eq. (\ref{eq:pi-def}), which mostly contribute to Landau damping term, are orthogonal to ${\bf q}$ (Ref.~\cite{Chubukov2003}).
 The full bosonic susceptibility at $\q_0 \ll \vf |\bf q|$ then becomes
  \begin{equation}
  \label{eq:sus-def_1}
  D(q) = \frac{\chi_0}{\xi^{-2} + \q^2
  + \gamma f^2({\hat q}')\frac{|q_0|}{\vf |\q|}  + q_0^2/c^2}.
\end{equation}
At frequencies well below $\omega_1$, the regular $q^2_0/c^2$ term is smaller than the Landau damping term and can be safely neglected.

Substituting Eq. (\ref{eq:sus-def_1}) into (\ref{eq:sigma-def}) and factorizing the momentum integration, we obtain 
\begin{align}
  \label{eq:se=1-loop}
  i\Sg(k) &= i \w_0^{1/3}|f(\hat{k})|^{4/3} k_0^{2/3} + i k_0\frac{\gb}{\ef}\lambda(\hat k,k_0).
\end{align}
Here,
\begin{align}
  \w_0 &= \left(\frac{\gb}{2\pi\sqrt{3}}\right)^{3} \frac{1}{\gamma\vf^2}~ = \frac{1}{24\pi^2\sqrt{3}}\frac{\gb}{N\ef},
\end{align}
and
\begin{widetext}
\begin{align}
  \label{eq:lambda-def}
  \lambda (\hat k, k_0) &=
  \frac{1}{2\pi^2} \int_0^{1} dx\int d\phi \left[\frac{f^2(\phi_k+ \frac{\phi}{2})|\phi|}{|\phi|^3+\frac{2f^2(\theta(\phi,\phi_k))\gamma|k_0||x+1|}{\ef\kf^2}} - \frac{f^2(\phi_k)|\phi|}{|\phi|^3+\frac{2f^2(\phi_k)|k_0|\gamma|x+1|}{\ef\kf^2}}\right]\nn\\
   &=
     \lambda_0(\hat k) + \frac{\gb}{\ef}
     \left|\frac{k_0}{\w_0}\right|^{1/3}\lambda_1(\hat k).
\end{align}
\end{widetext}
Here $\phi\{\cos\theta(\phi_k,\phi),\cos\theta(\phi_k,\phi)\}$ parameterizes the position of $\k + \p$ on the FS. $\lambda_0 (\hat k)$ and $\lambda_1 (\hat k)$ are some angle-dependent parameters of order one. Note that, in accordance with Eq. \eqref{eq:pi-1-loop}, we have $\hat\theta =\hat z \times \hat p \simeq \hat z \times \hat z \times \hat k = -\hat k$. The variation of
$\theta + \hat k$ just modifies the form of $\lambda_1$ somewhat. In Eq. \eqref{eq:lambda-def} we also assumed $f^2(\k) = f^2(-\k)$. Henceforth for simplicity we drop the variation of $\theta$ and simply replace $\hat\theta\to -\hat k \to \hat k$. Finally note that when $f=1$, $\lambda (\hat k, k_0)$ vanishes identically (at the level of the integrand in Eq. (\ref{eq:lambda-def}) above).

The validity of the factorization of momentum integration in Eq (\ref{eq:sigma-def}) is verified {\it a posteriori}.  Typical internal momenta and frequencies in the integrals are
\begin{align}
  \label{eq:char-mom-freq}
  \w &\sim \w_0,\quad k-k_F \sim q_\perp \sim \Sg \sim \w_0/\vf, \nonumber \\
  q_\parallel &\sim
   (\gamma \w_0/\vf)^{1/3} \sim q_\perp \left(\frac{N \ef}{\bar g}\right)
\end{align}
We see that, as long as $N \ef/{\bar g} \gg 1$, typical $q_\parallel$ are much larger than typical $q_\perp$ and $k-k_F$. This is the justification for the factorization of the momentum integration. One can also check that at these $\omega$ and $q$, vertex corrections are small in ${\bar g}/(N\ef)$ (Ref.~\cite{Kim1994}).
[To be exact, the $\lambda_1$ term has a contribution of order one from momenta of order $q_\parallel\sim q_\perp \sim \Sg$, which is formally beyond the justification of the momentum factorization (see Appendix \ref{sec:derivation-eqs-pi-sig}). However, because we are  not interested in the exact form of $\lambda_1$, we can safely neglect this contribution.]

\subsection{The vanishing of $\Pi$ for a constant form factor}
\label{sec:vanish-pi-const}
 The ladder series of vertex renormalizations for $f = 1$ have been  analyzed in Ref.  \onlinecite{Chubukov2005}.  The full vertex $\Gamma (k, q) \approx \Gamma (k_0, q_0)$, evaluated at $\q=0$ and general $\k$, obeys an integral equation, whose solution is
\begin{align}
  \label{eq:ward-eliashberg}
  i q_0 \Gamma(k_0; q_0) &= G^{-1}(k+q) - G^{-1}(k) \nn\\
                                   &= i\Sgt(k_0+q_0) -i\Sgt(k_0).
\end{align}
This coincides with the Ward identity for the density vertex~\cite{Peskin1995}.
Eq. \eqref{eq:ward-eliashberg} is equivalent to:
\begin{flalign}
  \label{eq:Gamma-sigma-gg}
  &\Gamma(k;q)G(k)G(k+q) = \nn\\
  &\qquad\qquad\qquad\frac{G(k)-G(k+q)}{i q_0}.
\end{flalign}
Plugging this into  Eq. \eqref{eq:pi-def_1} we find
\begin{equation}
  \label{eq:pi-ward}
  \Pi (\q=0, q_0) \propto q_0^{-1}\int d^3k
  \left[G(k+q) - G(k)\right].
\end{equation}
We recall that the integral $\int d^3k$ is
\begin{equation}
  \label{eq:cutoff-reg}
  \int d^3k = \frac{\kf}{\vf} \int_{-\infty}^{\infty}
  \frac{dk_0}{2\pi} \int^\Lambda d\epsilon_k \int_0^{2\pi} \frac{d\theta_k}{2\pi}
\end{equation}
where  $\Lambda$ is the upper energy cutoff of the low-energy model.   One can immediately check that $\int d^3 k G(k)$ is ultraviolet convergent.

The second term in  (\ref{eq:pi-ward}) can be transformed into the first term by shifting integration variable $k$ by external $q$. In general, such shift has to be taken with care because one also has to shift the upper limit of integration over $\epsilon_k$. In our case, however, the momentum component of $q$ is zero, and the shift only involves the frequency component, over which the integration holds in infinite limits. As a result,
\begin{equation}
\int d^3k G(k+q) =  \int d^3k  G(k)
\end{equation}
and, hence $\Pi (\q=0, q_0) =0$, as long as $q_0$ is finite.

Another way to obtain the same result is to write $G(k+q) - G(k) = i(\tilde \Sigma (k_0) - \tilde \Sigma (k_0+q_0)) G(k+q) G(k)$ such that
\begin{align}
  \label{eq:pi-ward_1}
  \Pi (\q=0, q_0) &\propto q_0^{-1}\int d k_0 \int d \epsilon_k \int d\theta_k\nn\\
  &\qquad\left[(\tilde \Sigma (k_0) - \tilde \Sigma (k_0+q_0)) G(k+q) G(k)\right],
  \end{align}
  and integrate in (\ref{eq:pi-ward_1}) first over fermionic dispersion and then over frequency.
 The integral  has two contributions: one comes from the range $\epsilon_k \sim k_0 \sim q_0$, where the poles in $G(k+q)$ and in $G(k)$ are in different half-planes of $\epsilon_k$, once we extend $\int d \epsilon_k$ onto a complex plane.  The second contribution comes from high energies $\epsilon_k \sim k_0 \sim \Lambda$. At such energies, $\Sigma (k_0) \ll k_0$, i.e., ${\tilde \Sigma} (k_0) \approx k_0$.
 Evaluating the two contributions, we find that they exactly cancel each other:
\begin{equation}
  \label{eq:low-energy-const}
  \Pi (q=0, q_0)_{low}  = \gamma\gb^{-1},
  \quad\Pi (q=0, q_0)_{high} = -\gamma\gb^{-1}.
\end{equation}
Thus, the $\Pi = 0$ result comes from an exact cancellation between low- and high- frequency terms. We may expect that for $f \neq 1$ the high frequency piece will remain essentially unchanged. However, the low frequency piece will get additional contributions from the variation of $f(\k)$ along the FS, leading to a nonzero $\Pi$.

\subsection{Calculating $\Gamma$ for $f \neq 1$}
\label{sec:calculating-gamma-f}

We now perform the same calculation for angle-dependent $f(\k)$. We express the vertex function $\Gamma (k, q)$ at $q = (0, q_0)$  as
\begin{equation}
  \label{eq:Gamma-dyson}
  \Gamma(k; q) = f(\hat{k})\left[1 + \delta\Gamma(k; q)\right].
\end{equation}
The ladder equation for $\delta\Gamma(k; q)$ is
\begin{widetext}
\begin{align}
  \label{eq:delta-Gamma-def}
f(\hat k)  \delta\Gamma(k; q) &=
                       \frac{\gb}{\chi_0\tpp^3}\int d^3p
                                f(\hat p)
                       [1+\delta\Gamma(p; q)]G(p+q)G(p)f^2\left(\frac{\k+\p}{2}\right)D(p-k).
\end{align}
To get an insight how $\delta\Gamma(k; q)$ should look like, consider first a simpler problem, namely the renormalization of the density vertex $\Gamma_0 (k, q) = 1 + \delta \Gamma_0 (k, q)$, still keeping angle-dependent $f$ in the interaction vertices. The density vertex correction $\delta\Gamma_0 (k, q)$ obeys
\begin{align}
  \label{eq:delta-Gamma-0-def}
  \delta\Gamma_0(k; q) &=
                         \frac{\gb }{\chi_0\tpp^3}\int d^3p [1+\delta\Gamma_0(p;q)]G(p+q)G(p)f^2\left(\frac{\k+\p}{2}\right)D(p-k).
\end{align}
We factorize the momentum integral and again employ Eq.  \eqref{eq:GG-kappa} to simplify the equation for $\delta\Gamma_0(k;q) \equiv \delta\Gamma_0(\hat k, k_0; \q=0, q_0)$  to
  \begin{align}
    \label{eq:dg-eliashberg-form}
   \delta\Gamma_0({\hat k},k_0,q_0) &= \frac{\gb}{\tpp^2 \vf }\int_{-q_0}^0 dp_0 \frac{1 + \delta \Gamma_0({\hat p},p_0,q_0)}{\Sgt(q+p) - \Sgt(p)}\int dp_\parallel\frac{f^2\left(\frac{\k+\p}{2}\right)|p_\parallel|}{|p_\parallel|^3+\gamma f^2(\hat{k})|p_0-k_0|/\vf}.
  \end{align}
\end{widetext}
To solve this equation, we note that the difference $\Sgt(k+q) -\Sgt(k)$ (which is a function of ${\hat k}$ and $k_0$) is expressed via the same integral as in the r.h.s. of (\ref{eq:dg-eliashberg-form}), namely
\begin{widetext}
  \begin{align}
    \label{eq:self-energy-eq}
    \Sgt(k+q)-\Sgt(k) &= q_0 + \frac{\gb}{\tpp^2\chi_0}\int d^3p [G(p+q)-G(p)]f^2\left(\frac{\k+\p}{2}\right)
                        D (p-k) \nn \\
                      &= q_0 \left(1 + \frac{\gb}{\tpp^2 \vf}\int dp_\parallel\frac{f^2\left(\frac{\k+\p}{2}\right)|p_\parallel|}{|p_\parallel|^3+\gamma f^2(\hat{k})|p_0-k_0|/\vf}\right).
  \end{align}
\end{widetext}
We then argue that
 \begin{equation}
  \label{eq:gamma-0-sol}
  \delta \Gamma_0(k; q) = \frac{\Sigma(k+q) - \Sigma(k)}{q_0} \equiv \frac{\Sgt(k+q)-\Sgt(k)}{q_0} -1
\end{equation}
is a solution of Eq. (\ref{eq:dg-eliashberg-form}).  One can verify this by just substituting Eq. \eqref{eq:gamma-0-sol} for $\delta \Gamma_0$ into the r.h.s. of (\ref{eq:dg-eliashberg-form}) and relating the integral over $p_{\parallel}$ in (\ref{eq:dg-eliashberg-form}) to $ \delta \Gamma_0(k;q)$ using Eq. (\ref{eq:self-energy-eq}).
The form of Eq. \eqref{eq:gamma-0-sol} is just that of a Ward identity for the density vertex, similarly to what was obtained for $f=1$, Eq. \eqref{eq:ward-eliashberg}. See Appendix \ref{sec:eliashb-vert-gamm} for details.

The Ward identity, Eq. \eqref{eq:gamma-0-sol},  is the expected result: it shows that the density-density polarization bubble vanishes at zero  incoming momentum and finite frequency, even for a system with fermion-fermion interaction in the nematic channel. To see this explicitly, we plug  (\ref{eq:gamma-0-sol}) into the formula for density-density polarization
\begin{equation}
\label{eq:pi-def_2}
  \Pi^{\rho} (q) =
  N \gb
  \int \frac{d^{3}k}{\tpp^3}\Gamma_0 (k;q) G(k)G(k+q),
  \end{equation}
 approximate $\int d^{3}k/\tpp^3$ by $\int d k_0 \int^\Lambda d \epsilon_k$ using Eq. (\ref{eq:cutoff-reg})    and integrate first over $\epsilon_k$ and then over $k_0$. As we discussed earlier, the integral has high-energy and low-energy contributions. For the high-energy contribution, the self-energy and vertex correction can be neglected, while for the low-energy contribution both are relevant.  Evaluating the integrals, we obtain
\begin{align}
  \label{eq:low-energy-const-f}
  &\Pi^{\rho} (\q=0, q_0)_{low}  =  \gamma \gb^{-1} \langle f^2 \rangle, \nn\\
  &\Pi^{\rho} (\q=0, q_0)_{high} = - \gamma \gb^{-1} \langle f^2 \rangle.
\end{align}
We recall that
\begin{equation}
  \label{eq:angular-avg-def}
  \langle f^2(\theta) \rangle = \int \frac{d\theta}{\tp}f^2(\theta).\nn
\end{equation}
The two contributions then cancel out for any $f(\hat k)$, i.e., the density-density polarization bubble $ \Pi^{\rho} (\q=0, q_0)$ vanishes, as it should, for arbitrary interaction between fermions that conserves the total number of particles.

We now use this result to analyze the integral equation (\ref{eq:delta-Gamma-def})  for
the correction to the full vertex, $\delta \Gamma (k,q)$.
The leading contribution to the renormalization of $\delta \Gamma (k, q)$ at each order comes from small momentum transfer $\k-\p$.
 It is therefore tempting to just replace
$f(\hat p)$ in the r.h.s. of (\ref{eq:delta-Gamma-def})
 with $f(\hat k)$.
However, in that case
we would obtain the same equation as for $\delta \Gamma_0 (k, p)$, i.e., within this approximation, $\delta \Gamma (k, q)$  would be equal to $\delta \Gamma_0 (k, q)$, and the effects of non-conservation of the order parameter would not show up in  the polarization operator.  To detect the effects due to non-conservation, we need to go beyond approximating  $f(\hat p)$  by  $f(\hat k)$, i.e., we need to include subleading terms, which account for the fact that $f(\hat p)$ is not identical to $f(\hat k)$. It is this difference that makes $\Pi (\q =0, q_0)$ finite, as we will see.

To single out the contribution which is sensitive to the variation between $f$ at internal and external momentum in the vertex correction diagrams, we make an ansatz
\begin{equation}
  \label{eq:dg-1-def}
 1 +\delta\Gamma (k, q) =  \frac{\Sgt(k+q)-\Sgt(k)}{q_0(1+\mu)}  \equiv  \frac{1 + \delta \Gamma_0 (k, q)}{1+\mu ({\hat k}, k_0,q_0)},
\end{equation}
where $\mu ({\hat k}, k_0,q_0)$ is the term that accounts for the difference in $f$ that we are interested in.
Plugging the ansatz into Eq. (\ref{eq:Gamma-dyson}) we obtain
 \begin{equation}
  \label{eq:dg-1-ansatz}
  \Gamma({\hat k},k_0,q_0) =  f (\hat k) \frac{1 + \delta \Gamma_0({\hat k},k_0,q_0)}{1+\mu({\hat k}, k_0,q_0)}
\end{equation}
We assume and then verify that $\mu$ is small and expand in $\mu$.
By direct comparison with Eq. \eqref{eq:delta-Gamma-def} we then find
\begin{widetext}
\begin{align}
  \label{eq:mu-expr}
  \mu ({\hat k}, k_0,q_0) f(\hat k) &= \frac{\gb }{\chi_0\tpp^2\vf\kf}\int_{-q_0}^{0} \frac{dp_0}{q_0}\int \kf dp_\parallel [f(\hat k)- f(\hat p)]f^2\left(\frac{\k+\p}{2}\right)D(p-k)  + O(\mu^2)\\
                &= \frac{\gb}{\ef}\left[\mu_0(\hat k) + \frac{\gb}{\ef}
                  \left|\frac{q_0}{\w_0}\right|^{1/3}\mu_1\left(\hat k, \frac{k_0}{q_0}\right)\right], \label{eq:mu-def}
\end{align}
\end{widetext}
where $\mu_0 (\hat k)$ and $\mu_1\left(\hat k, \frac{k_0}{q_0}\right)$ are dimensionless functions with $O(1)$ dependence on parameters. Substituting this $\mu ({\hat k}, k_0,q_0)$ into Eq. (\ref{eq:dg-1-ansatz}) for $\Gamma ({\hat k},k_0,q_0)$, plugging the vertex into the expression for $\Pi (\q=0, q_0)$, and evaluating the integral by integrating over dispersion first and then over frequency, we obtain
v\begin{align}
  \label{eq:pi-full-result}
  \gb\Pi  (\q=0, q_0) &= \gamma \int_0^{q_0}\frac{dk_0}{q_0} \int \frac{d\phi}{\tp}\frac{f^2(\phi)}{1+\mu(\phi,q_0,k_0/q_0)}-\gamma\langle f^2\rangle \nn \\
  &= -\gamma \int_0^{q_0}\frac{dk_0}{q_0} \left\langle\mu f^2\right\rangle + O(\mu^2) \nn \\
                      &\simeq N k^2_F \left(\frac{\bar g}{\ef}\right)^2
                        \left(A + C  \left(N \frac{\bar g |q_0|}{\epsilon^2_F}\right)^{1/3}\right)
\end{align}
where
\begin{align}
  \label{eq:pi-full-result_1}
  A &= -\langle \mu_0 f^2\rangle  \nn \\
  C &= - \langle {\bar \mu}_1 |f|^{8/3}\rangle
\end{align}
and ${\bar \mu}_1(\hat k) = \int_0^1 dx \mu_1(\hat k, x)$. This $\Pi  (\q=0, q_0)$ has the same form as Eq. (\ref{eq:pi-bare-wrong}) that we obtained in the leading order in the expansion in bosonic propagators. Moreover, the prefactors $A$ and $C$ in (\ref{eq:pi-bare-wrong}) and in (\ref{eq:pi-full-result}) are exactly the same (see Appendices \ref{sec:deriv-results-sec-bare}+\ref{sec:eliashb-vert-gamm} for more detail).

We now see that the functional form of the full $ \Pi  (\q=0, q_0)$ does not change between $q_0 >\w_0$ and $q_0 < \w_0$. The reason for this is that $\w_0$ is the scale where NFL behavior sets in, leading to nonanalytic self energy and singular vertex corrections. However, these corrections are local in space, and so for small momentum transfer $\vf |\q| \ll \ef$ the leading-order dependence of $\Pi$ on vertex corrections is the same as for the density-density polarization. The small nonzero polarization comes from virtual processes with large momentum transfer that are subleading to the nonanalytic part and do not depend on it.

\subsection{Deriving Eq. \eqref{eq:pi-full-result} by analyzing ladder contributions rung-by-rung}
\label{sec:deriving-rung-rung}

\begin{figure*}
  \centering
  \subfloat{
    \includegraphics[width=0.33\textwidth,clip,trim=120 500 150 120]{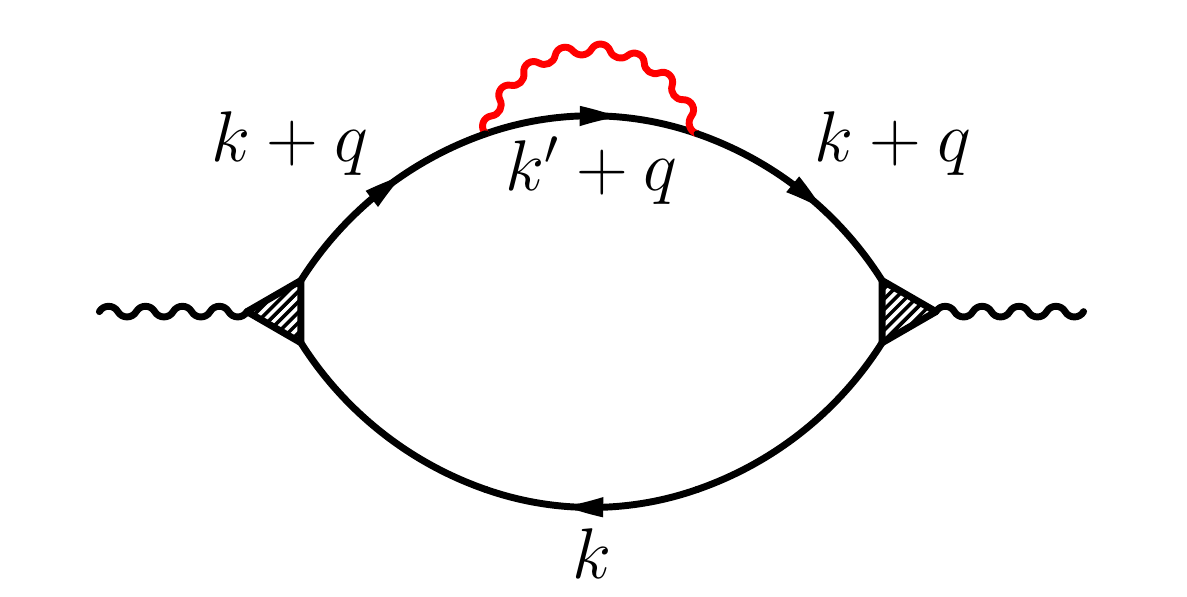}}
  \subfloat{
    \includegraphics[width=0.33\textwidth,clip,trim=120 500 150 120]{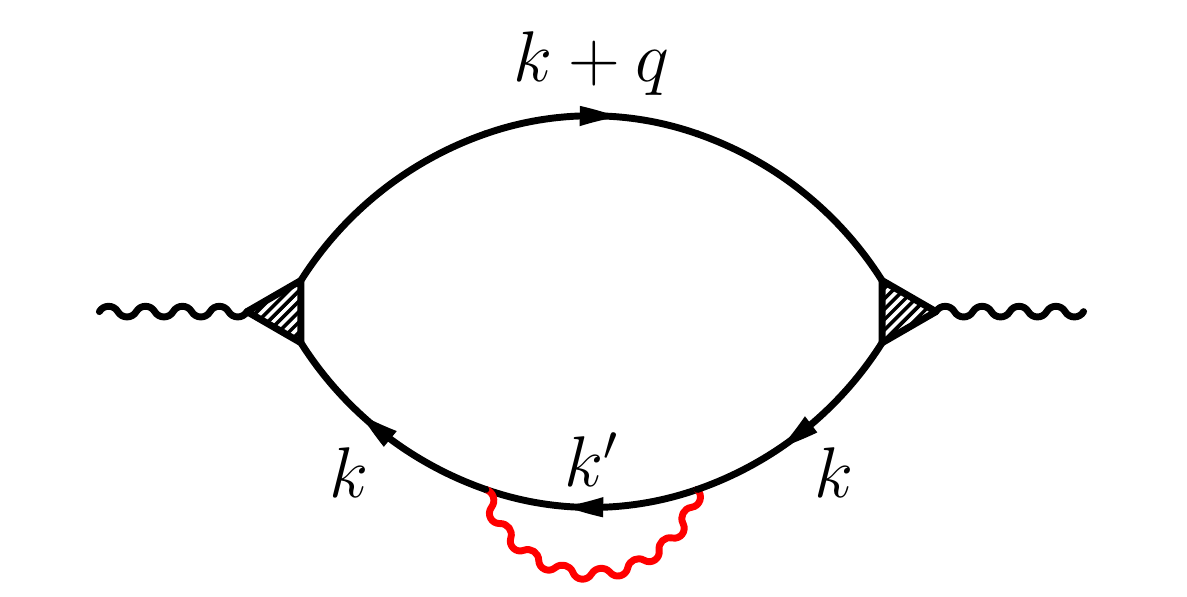}}
  \subfloat{
    \includegraphics[width=0.33\textwidth,clip,trim=120 500 150 120]{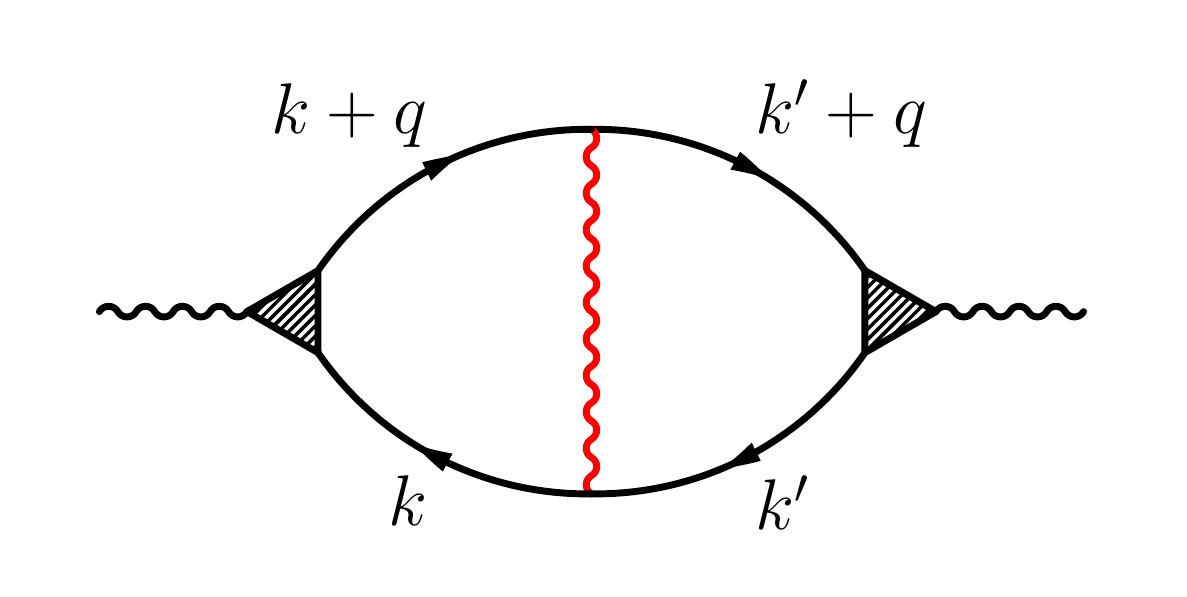}}
  \caption{The contributions to the polarization with vertex corrections taken into account. In order to obtain the low energy behavior, it is necessary to compute the full bubble (Fig. \ref{fig:pol-bubble}) within a ladder approximation. For a constant form factor, the total polarization is zero. In order to compute the correction coming from the form factor, we insert bosonic propagators with form factor vertices in each cross section of each bubble diagram (red wiggly lines). On all other cross sections we treat the form factor as constant. The result is a set of diagrams with density vertices $\Gamma_0$ on the sides. The expression for their sum is given in Eq. \eqref{eq:Id-correct-1}.}
  \label{fig:full-three-diags}
\end{figure*}

In the derivation of Eq. (\ref{eq:pi-full-result}) we explored the fact that the nematic vertex has momentum-dependent form-factor $f(\k)$, which varies a bit between external and internal momenta once we include vertex renormalizations. We now derive the same result in a different manner. Namely, we write the full polarization bubble, consisting of a sum of ladder contributions, as $\Pi = \sum_n \Pi_n$, where
\begin{align}
  \label{eq:pi-n-def}
  \Pi_n &\propto \int d^3k_1 f(\k_1) G(k_1+q)G(k_1) \times \nn \\
  &\qquad\hat{O}(k_2) \cdots \hat{O}(k_n)f(\k_n),
\end{align}
and
\begin{align}
  \label{eq:O-def}
  \hat O(k_j) &= \int d^3k_j f^2\left(\frac{k_{j-1}+k_j}{2}\right)\times\nn\\
  &\qquad D(k_j-k_{j-1})G(k_j+q)G(k_j).
\end{align}
Each $\hat O(k_j)$ represents a ``rung'', which consists of two fermionic propagators, and the effective interaction
\begin{equation}
  \label{eq:eff-int}
  U(k_j,k_{j-1}) = f^2\left((k_j+k_{j-1})/2\right) D(k_j-k_{j-1}).
\end{equation}
The fermionic propagator also contains $U(k_j,k_{j-1})$ via the self-energy, hence in each rung there are three "sources" for the dependence on  $f(\k)$. We assume  and then verify that the correction to the polarization, coming from variation of the form factor, is small. In this situation we may obtain $\Pi$ by separately calculating the contribution from each term (in each segment). Furthermore, within each rung  we select \emph{one} of its three $U$'s, where we allow $f(\k)$ to vary, and hold $f(\k)$ in all other $U$'s constant. We then repeat the procedure for the other two $U$'s in $\hat O(k_j)$. Finally we sum up contributions from all $j$ in all $\Pi_n$.

To see how this works in practice, it is convenient to switch to symmetrized variables,
\begin{equation}
  \label{eq:sym-vars}
  {\bar k}_j = \frac{k_j + k_{j-1}}{2},\kappa_j = k_j - k_{j-1},
\end{equation}
so that by construction,
\begin{equation}
  \label{eq:sym-vars-2}
  k_0 = \bar k_1 - \frac{\kappa_1}{2},\quad k_j = \bar k_1 + \frac{\kappa_1}{2} +\kappa_2 + \cdots + \kappa_j
\end{equation}
For simplicity, let's consider the situation where we may expand the form-factors in the variables $\kappa_j$ to second order. Next, also for simplicity, let us consider the situation where we hold all the $U$'s from the fermionic propagators constant and expand only in the $U$'s which constitute vertex corrections in some bubble diagram $\Pi_n$. It is easy to see that we will get a series of terms proportional to $\kappa_a^2,\kappa_b^2$ and to $\kappa_a\kappa_b$, where $a \neq b$ are two segments in $\Pi_n$. However, $\kappa_a,\kappa_b$ are independent angular variables, and so upon integration, all cross terms vanish, leaving only those terms that depend on a \emph{single} segment variable $\kappa_a$ or $\kappa_b$, expanded to second order. Thus, to obtain all contributions dependent on the variable $\kappa_j$ in $\Pi_n$, we may replace all form factor terms for $i < j$ with $f(\bar k_1)$, and all those for $i > j$ with $f(\bar k_1 + \kappa_j)$. We repeat this process for each segment, and add them all up. It is readily verified that when we also consider self-energy corrections, the story does not change. In fact, it is possible to identify precisely which terms in the self-energy and vertex corrections cancel out. (This can be done by properly symmetrizing Eq. \eqref{eq:Fkk} which appears later in this section.)

In practice, we can do all the summations at once by calculating the three diagrams of Fig. \ref{fig:full-three-diags}. In each diagram, the effective interaction marked in red is allowed to vary, and all others are held constant. This is done by replacing the side vertices with $f(\hat k) \Gamma_0$ and for the fermionic self-energy using only the first term in Eq. (\ref{eq:se=1-loop}). (In the perscription we just gave it is not immediately clear why we are getting a \emph{correction} coming from the form-factor variation. To see this it is enough to try and calculate the three diagrams of Fig. \ref{fig:full-three-diags} without letting the form-factor vary within the red lines. It is readily verified that in that case the three diagrams sum to zero.)

We carry out the procedure we just outlined, collect contributions from the three diagrams, and obtain (see Appendix \ref{sec:deriv-results-sec} for details),
\begin{widetext}
\begin{align}
  \label{eq:Id-correct-1}
\Pi (\q=0, q_0) = N \frac{\gb}{\chi_0} \frac{1}{\tpp^6}\int d^3k d^3k' \Gamma_0 (k, q)
G(k)G(k+q)D(k-k')\mathcal{F}(\k;\k')G(k')G(k'+q) \Gamma_0(k', q).
\end{align}
where
\begin{equation}
  \label{eq:Fkk}
  \mathcal{F}(\k;\k+\p) = f(\k)f^2(\k+\p/2)[f(\k+\p) - f(\k)] = f_2(\hat{k}) \left(\frac{\p\times\hat{k}}{\kf}\right)^2 + \cdots,\qquad f_2 = f^2 f'^2 +\frac{1}{2}f^3 f''.
\end{equation}
\end{widetext}
and in the four fermionic $G$ the self-energy is given by the first term in Eq. (\ref{eq:se=1-loop}). Integrating over two fermionic dispersions and one frequency, we obtain after some algebra
\begin{align}
  \label{eq:Pi-full-2nd}
  \gb\Pi (\q=0, q_0) &= N k^2_F \left(\frac{\bar g}{\ef}\right)^2\int_0^{q_0}\frac{dk_0}{q_0}\int \frac{d\phi}{\tp}\int \frac{d\theta}{\tp} \times \nn \\
  &\qquad\qquad\qquad\frac{\mathcal{F}(\phi,\phi+\theta)|\theta|}{|\theta|^3 +\frac{ f^2(\phi)\gamma|k_0+q_0|}{\vf\kf^3}}
\end{align}
Evaluating the angular integrals and the integral over $k_0$ we indeed reproduce Eq. \eqref{eq:pi-full-result}.

The derivation we just gave illuminates the different roles played by processes with small momentum transfer and with large momentum transfer.  The scattering processes with  small momentum transfer renormalize the vertices and fermionic self-energies, but the renormalization is the same as if the form-factor was equal to one. The presence of the form-factor only gives rise to multiplication factors with $f$ at the same momentum at which we,
e.g., compute the self-energy.  These low-energy scattering processes do not sample enough of the FS to be aware of the form factor variation. The nonvanishing polarization comes from the processes in which a fermion scatters all along the FS, i.e., a characteristic scattering momentum is of order $k_F$. Note, however, that this separation only explains the frequency independent piece in $\Pi (\q=0, q_0)$. The frequency dependent $q^{1/3}_0$ term comes from small momentum scattering, but indeed it also originates from the variation between $f(\k)$  at the beginning and the end of the scattering process.

The consideration based on a selection of a segment, where the contribution comes from a range of momentum transfers, different from those in other segments, is similar to  diagrammatic derivation of the FL formula for the static susceptibility \cite{Chubukov2014a,Finkelstein2010}. It is also similar to the derivation of a conductivity in a metallic system in terms of transport lifetime, either due to impurity scattering \cite{Abrikosov1975}, or due to electron-electron interaction, particularly near a QCP \cite{Maslov2017}.

A comment is in order: in our analysis we ignored the existence of ``cold-spots'' - the points on the Fermi surface where the form-factor has nodes (these are along the directions $\pi/4+n \pi/2$ for a d-wave form factor). At the cold spots the form factor vanishes, and hence the self-energy and Landau damping vanish, but at different rates. It can be shown that to get fermionic self-energy and the Landau damping near the cold spots, one must go beyond the Eliashberg approximation \cite{AvrahamKleinUnp2017}. However, these effects are not significant for the computation of the polarization bubble at zero momentum and finite frequency as this polarization comes from processes around the entire FS, and the cold spot contributions are negligible.

We can refine the estimate for $\Pi$ a bit by expanding $\mathcal F$ as in Eq. \eqref{eq:Fkk}.
Then, for  $f(x) = \cos{\ell x}$, we find
\begin{equation}
  \label{eq:f2-ell}
  \langle f_2^\ell \rangle = - \frac{\ell^2}{16}
\end{equation}
Applying this result to the  d-wave case, when $\ell=2$, we find
\begin{equation}
  \label{eq:f2-ell-2}
  \langle f_2^{nem} \rangle = - \frac{1}{4}
\end{equation}
Because the constant term in $\Pi (\q=0, q_0)$ is proportional to $ \langle f_2^{nem} \rangle$ and the constant and  the $|q_0|^{1/3}$ terms have opposite signs (this immediately follows from  (\ref{eq:Pi-full-2nd}),  in Eq. (\ref{eq:pi-full-result}) $A <0$  and $B >0$. This is an expected result because with our sign conventions the nematic susceptibility in real frequencies has an imaginary part $D'' (\q=0,\W) \sim -\Pi'' (\q=0,\W)$. Thus  $D^{''} (\Omega) \sim  - B \Im (-i \Omega)^{1/3}$ has the same sign as $\Omega$, as it should, by causality principle.

It is natural to ask what is the contribution from terms in which the gradient of $f(\k)$ is kept in more than one segment.
   In the diagrammatic computation of the spin susceptibility in a Fermi liquid, the diagrams with one ``special'' segment (where the integration is confined to infinitesimally small vicinity of  the FS) gives $m^*/m$, while diagrams with two, three, etc. such segments yield a geometric series $(-1)^n F^n_l$, in powers of the Landau parameter $F_l$. The sum of such terms gives the $1/(1+  F_l)$ term in $\chi_l$ \cite{Finkelstein2010,Chubukov2014a}.  In our case, we expect that a similar computation will  yield a series of $q^{1/3}_0$ terms, which likely do not lead to any new physics.

\section{Nematic susceptibility and Raman response}
\label{sec:Raman}

\begin{figure*}
  \centering
  \subfloat[\label{fig:susc-a}]{
    \includegraphics[width=0.49\hsize,clip, trim = 0 280 0 150]{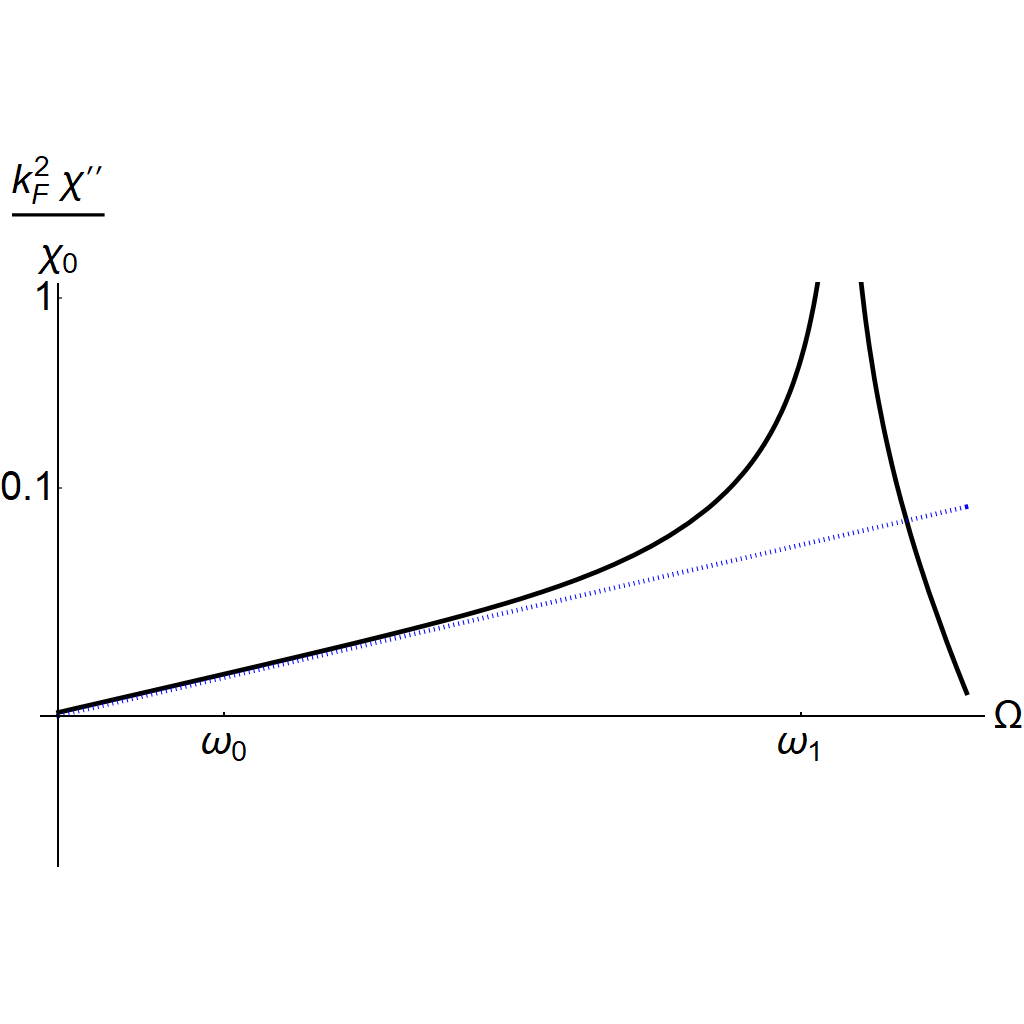}}
  \subfloat[\label{fig:susc-b}]{
    \includegraphics[width=0.49\hsize,clip, trim = 0 280 0 150]{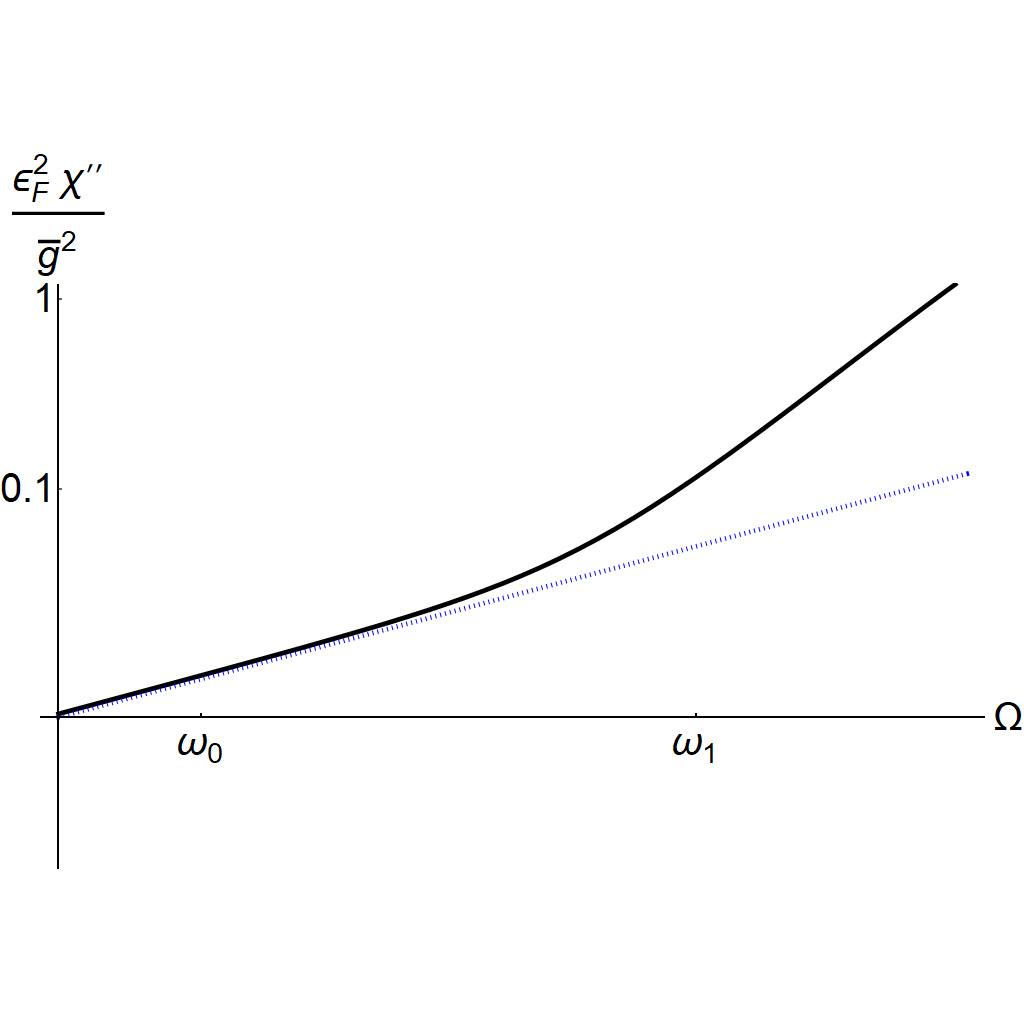}}
  \caption{Plots of the nonzero dynamic response for a nematic order parameter(with $N=1$). (a) Dynamical response $D''(\q=0,\W)$ for independent bosons (such as phonons near a structural transition)  coupled to itinerant fermions, . (b) Dynamical response $\chi''(\q=0,\W)$ for a system of interacting fermions only, such as a $d-$wave Pomeranchuk transition. In both cases the low frequency response is $\sim \W^{1/3}$ (dotted reference line), in the region $\W \ll \w_1$. The behavior does not change across the onset frequency for NFL behavior $\w_0$. At higher frequencies $\W > \w_1$, the response becomes linear. For a transition involving independent bosons, the linear behavior is masked by a peak near $\Omega=\omega_1$.}
  \label{fig:susc-comp}
\end{figure*}

The uniform dynamic susceptibility of the nematic $\phi$ field at a QCP  is related to $\Pi (\q=0,q_0)$ by
\begin{equation}
D(\q=0,q_0) = \frac{\chi_0}{\xi^{-2}_0 + q^2_0/c^2
  +  \gb \Pi (\q=0,q_0)}
\label{ch_n1}
\end{equation}
where $\xi_0^{-2} = \gamma \langle f^2\rangle \sim N \kf^2 (\gb/\ef)$ (because $\xi^{-2} = \xi^{-2}_0 - \gamma \langle f^2\rangle =0$). The functional form of $D(\q=0,\W_m)$ can be directly probed in Monte-Carlo studies.  Recent studies \cite{Lederer2017,Schattner2016} have used the same model as ours -- a scalar bosonic field undergoes an Ising-nematic transition, and the susceptibility of $\phi$ field  gets modified by the  minimal coupling to fermions.

The imaginary part of nematic susceptibility in real frequencies, $D^{''} (\q=0, \Omega)$ can be directly measured in Raman experiments\cite{Devereaux2007,Gallais2013,Gallais2016,Thorsmolle2016}.
To obtain the frequency dependence of $D^{''} (\q=0, \Omega)$ we note that the constant term in $\gb\Pi (\q=0,q_0)$ is of order $N k^2_F ({\bar g}/\ef)^2$, i.e., is small relative to $\xi^{-2}_0$.  The dynamic term is even smaller, but it is non-analytic in frequency and hence it has a non-zero imaginary component. Converting to real frequencies ($q_0 \to -i \Omega$)  and expanding in small $\gb\Pi (\q=0,\Omega)/\xi^{-2}_0$, we obtain
\begin{equation}
  D^{''}(\q=0,\Omega) \approx
  -
  \gb\Pi^{''} (\q=0,\Omega) \xi^4_0  \frac{\chi_0}{(1 - (\Omega \xi_0/c)^2)^2}
\label{ch_n2}
\end{equation}
The frequency dependence in the denominator becomes relevant at $\Omega
 \approx c/\xi_0 \sim \omega_1 (\vf/c)^{1/2} \sim \omega_1$.   At much smaller frequencies,
\begin{equation}
  D^{''}(\q=0,\Omega) \approx
  -
  \chi_0  \gb\Pi^{''} (\q=0,\Omega) \xi^4_0 \propto  \frac{\chi_0}{N k^2_F} \left(\frac{N\gb \Omega}{\ef^2}\right)^{1/3}
\label{ch_n3}
\end{equation}
At much larger frequencies,
\begin{equation}
D^{''}(\q=0,\W) \propto   N\frac{\chi_0}{k^2_F} \frac{\gb\ef^2}{\Omega^3}  \left(\frac{ c}{\vf}\right)^4.
\label{ch_n4}
\end{equation}
In between, there is a resonance at $\W  \approx c/\xi_0  \sim \w_1$, as seen in the peak in the first panel of Fig. \ref{fig:susc-comp}.
We note that the Raman signal will also include a response from the fermions themselves. However, at $\q = 0$, this signal will scale as $\Pi D^{-1} \sim (\gb/\ef)^2$, so it will be small.

One potential class of materials to which our results can be applied, are Fe-based systems, in particular FeSe doped by S, for which a nematic QCP separate from a magnetic QCP has been detected, and this QCP is only slightly masked by superconductivity \cite{Coldea2017}.
However, for applications to Fe-based systems our analysis likely has to be modified.  One obvious reason is the multi-band structure of Fe-based systems and the rather small value of $\ef$. But there is also another, more fundamental reason, related to the mechanism for nematicity.  In our approach we assumed that a scalar field $\phi$ acquires a nematic order independent on fermions.  For Fe-based systems, this would imply that nematicity develops via a structural transition,  i.e., that the order parameter field $\phi$ is a phonon field. In this case, fermions do modify the susceptibility of the $\phi$ field, but the transition itself happens even if the coupling ${\bar g}$ vanishes.  In Fe-based systems, nematicity is most likely of electronic origin and is either a transition to a composite spin order, or a Pomeranchuk  instability of the Fermi surface.  The order parameter for a Pomeranchuk instability couples in a minimal way to $d$-wave fermionic density, like in our model. The difference is that in a Pomeranchuk case the primary nematic field $\phi$ is by itself bilinear in fermions and describes $d$-wave collective charge fluctuations in a fermionic system.  As a consequence, the bosonic susceptibility is actually the $d$-wave charge susceptibility of interacting fermions.

The model of interacting fermions near a $d$-wave Pomeranchuk instability is  similar, but not identical to the model of a critical $\phi$ field coupled to fermions, and the distinction becomes pronounced at $\q=0$ and finite $\W_m$.  Indeed, the $d$-wave susceptibility of interacting fermions can, at least qualitatively, be described within RPA.  We label this susceptibility as $\chi (q, \W_m)$ to distinguish it from $D(q, \W_m)$. We have
\begin{equation}
  \label{eq:RPA-susc}
  \chi(\q, \W_m) = - \frac{\Pi (\q, \W_m)}{1 - U_d \Pi(\q, \W_m)}
\end{equation}
where $U_d < 0$ is an attractive fermion-fermion interaction in a d-wave channel. At low frequencies, when $\W_m \ll \vf |\q|$, $\Pi (\q, \W_m) = -a +  \gamma \Omega/(\vf q) + q^2 + \Omega^2_m/c^2 + ...$, where $a >0$, $\gamma$, and $c$ are microscopic parameters.
Because the constant $a$ term is the largest,  $\Pi (\q, \W_m)$ in the numerator in (\ref{eq:RPA-susc}) can be approximated by a constant.  In the denominator,
$1 - U_d \Pi (0,0)$ is set to be proportional to $\xi^{-2}$. Introducing $\chi_0$ to get $\chi (\q,0) = \chi_0/|{\bf q}|^2$ at large enough momentum,  we obtain at $\W_m \ll \vf |\q|$,
\begin{equation}
  \label{eq:RPA-susc_1}
  \chi(\q, \W_m) = \frac{\chi_0}{\xi^{-2} + q^2 + \W^2_m/c^2 + \gamma |\Omega_m|/(\vf |\q|)}
\end{equation}
This susceptibility has the same form as  $D(q,\W_m)$ in Eq. (\ref{ch999}).
\footnote{The presence of the $\Omega^2_m$ term in the bare susceptibility in
  ``fermion-only'' is actually questionable as for a conserved order parameter $\chi (\q=0, \Omega_m)$ must vanish for all $\Omega_m$, and for a non-
  conserved order parameter we will argue that the effects due to non-conservation are small at weak coupling.  Also, it has been argued recently~\cite{Maslov2017} that the prefactor for the $q^2$ term in the denominator of $\chi_0 (\q, \Omega_m)$  may actually be quite small, at least in some microscopic ``fermion-only'' models.}
However, in the  opposite limit  $\W_m \gg \vf |\q|$ that we are interested in, $\Pi (\q,\W_m)$ is small, and, to a good accuracy, we just have $\chi (q,\W_m) =-\Pi (q,\W_m)$. Then $\chi^{''} (\q=0, \Omega) = -\Pi^{''} (\q=0, \Omega)$.  Using our results for $\Pi$, we then obtain
\begin{align}
  \chi^{''} \propto & N\left(\frac{\gb}{\ef}\right)^2
                      \begin{cases} B\frac{\W}{\ef},\quad &\omega_1\ll\W\ll\ef\\
                        \frac{1}{2} C \left(\frac{N\gb\W}{\ef^2}\right)^{1/3},\quad&\W\ll\omega_1
                      \end{cases}
\end{align}
where $ B, C$ are the dimensionless constants of order one, previously discussed in the text. Fig. \ref{fig:susc} depicts the susceptibility over a range of frequencies, and Fig. \ref{fig:susc-comp} shows a comparison between the susceptibilities of independent vs. fully fermionic nematic orders.

Strong, near-critical nematic fluctuations have been found to be ubiquitous among Fe-based superconductors near optimal doping\cite{Thorsmolle2016,Gallais2016,Kuo2016},
and many of these materials have a substantially two-dimensional electronic structure.
However, the multi-band  electronic structure of these systems, as  well as the blurring of the Fermi surface due to thermal and disorder effects, have been found to play an important role in the Raman response of these materials\cite{Gallais2016}. We have not taken such effects into account in this work, so our predictions must come with additional conditions for their validity.  Also, our scaling forms apply to frequencies well above both $T$, but still low enough that contributions from optical phonons, among other excitations, can be neglected.

\section{Conclusions}
\label{sec:discussion}

In this work we computed the  polarization bubble at zero momentum and finite frequency, $\Pi (\q=0, \W_m)$  for fermions at a QCP towards $d_{x^2-y^2}$ nematic order.  The corresponding order parameter is not a conserved quantity, hence there is no conservation law that would require
$\Pi (\q=0, \W_m)$  to vanish.  We indeed found that  $\Pi (\q=0, \W_m)$  is non-zero, with a constant as the leading term.  The dynamic part $\Pi (\q=0, \W_m) - \Pi(\q=0, \W_m \to 0)$ is proportional to $ |\W_m|$ at high frequencies, crossing over to $|\W_m|^{1/3}$ at lower frequencies.

Though our analysis relied on weak coupling to control the calculations, we consider it plausible that similar phenomenology may prevail in real materials, where the coupling is of order one. In any case, proximity to a QCP with a nonconserved order parameter must on general grounds lead to nontrivial dynamics at zero momentum transfer. This regime is readily detectable in experiments such as Raman scattering, but has not been thoroughly explored in the theoretical literature. We hope our work provides motivation for its further study.

\begin{acknowledgments}
  We thank  M. Schuett, R. Fernandes, S. Kivelson, and M. Punk for stimulating discussions. This work was supported  by the NSF DMR-1523036 (AK and AC). SL and DC are supported by a postdoctoral fellowship from the Gordon and Betty Moore Foundation, under the EPiQS initiative, Grant GBMF-4303, at MIT.
\end{acknowledgments}
\bibliography{QCP,QCP_AC,NZS}

\clearpage
\onecolumngrid
\appendix

\section{Computational details of the perturbative evaluation of $\Pi^{(1)}(q)$}
\label{sec:deriv-results-sec-bare}

In this appendix we derive the results of sec. \ref{sec:incons-bare-pert} and the first part of section \ref{sec:breakd-eliashb-theor}. Namely, we calculate the diagrams of Fig. \ref{fig:2-loop}, and show that incorporating the $\w^{2/3}$ self-energy in the fermionic propagator gives a response even for a constant form factor. Our starting point is Eq. \eqref{eq:pi-2-bare} for the three contributions from the diagrams of Fig. \ref{fig:2-loop}. The calculation has three steps: first we identify contributions that contribute to the static part $\Pi(\q, q_0=0)$ and the dynamic part $\Pi(\q=0,q_0)$. Next we evaluate these contributions, including the Landau damping term in the bosonic propagator, but take the free propagator for fermions. We obtain Eq. \eqref{eq:pi-bare-wrong}. Then, we reevaluate the dynamic part, taking into account the self energy $\w^{2/3}$ term, and show that the polarization is nonzero even for $f=1$.

\subsection{Perturbative evaluation of $\Pi$ for $q_0 \gg \w_0$}
\label{sec:pert-eval-pi}

We start by splitting the two self energy contributions using the identity Eq.~\eqref{eq:GG-kappa},
\begin{align}
  \label{eq:app-I-pm-1}
  I_\pm &= \frac{N\gb}{\chi_0\tpp^6}\int d^3k d^3p ~ G^2(k)G(k+p)G(k \pm q) D(p) f^2(\k \pm \q/2)f^2(\k+\p/2) \nn \\
  &= \frac{N\gb}{\chi_0\tpp^6}\int d^3k d^3p ~ G(k)G(k+p)\mathcal{K}(k \pm q, k)[G(k) - G(k \pm q)] D(p) f^2(\k \pm \q/2)f^2(\k+\p/2)
\end{align}
The double Green's function $G(k)$ has no counterpart in the vertex correction and should be unrelated to the Ward identity, since it can't be canceled out by the vertex part. It provides a static term $\propto |\q|^2$,
\begin{align}
  \label{eq:app-I-s-1}
  I_s &=  \frac{N\gb}{\chi_0\tpp^6} \int d^3k d^3p ~ G^2(k) G(k+p) D(p) f^2(\k+\p/2) \times \nn \\
      &\qquad\qquad\qquad \times \left[\mathcal{K}(k+q,k)f^2(\k+\q/2)+\mathcal{K}(k-q,k)f^2(\k-\q/2)\right].
\end{align}
Evaluating for free fermions using Eq. \eqref{eq:kappa-def} for $\mathcal{K}$ we get,
\begin{align}
  \label{eq:I-s-2}
  I_s &=  \frac{N\gb}{\chi_0\tpp^6} \int d^3k d^3p ~ G^2(k) G(k+p) D(p) f^2(\k+\p/2) \frac{f^2(\k+\q/2)-f^2(\k-\q/2)}{i q_0 - \vf\hat{k}\cdot\q} \nn\\
      &= \frac{\gb\kf}{\tpp^5\vf}\int d\theta \frac{i p_0 dp_0}{(ip_0 - \vf p\cos\phi)^2}\frac{p^2 dp d\phi}{p^3+\gamma f^2(\phi)|p_0|/\vf}f^2(\theta+p\sin\phi/2)\frac{f^2(\theta+q\sin\theta/2)-f^2(\theta-q\sin\theta/2)}{i q_0 - \vf q \cos\theta}
\end{align}
where $\cos\theta = \hat{q}\cdot \hat{k}, \cos\phi = \hat{k}\cdot\hat{p}$, and we neglect the $p_0^2/c^2$ term in $D$ out of anticipation that its contribution can be neglected. The contribution from the region $\cos\phi \sim 0$ is zero because of the double pole. Thus, the static part comes from processes beyond the Eliashberg regime, i.e. by taking $q_0\to 0$ in the first denominator of Eq. \eqref{eq:I-s-2}. Taking the $q_0\to 0$ limit we get
\begin{align}
  I_s  &\sim \frac{\gb}{\ef^2\tpp^3}\frac{\w_0^{4/3}}{(\gamma\vf^2)^{2/3}}\langle f_3 \rangle q^2
\end{align}
where $\langle f_3 \rangle = \langle f^2(f^2)''' \rangle$. This correction is small in $1/\gamma$ and can be safely neglected.

The leftovers from Eq. \eqref{eq:app-I-pm-1} along with the vertex correction yield the dynamic part. Adding up the SE contributions we find
\begin{flalign}
  \label{eq:I-d-1}
  I_d^{SE} &= - \frac{N\gb}{\chi_0\tpp^6} \int d^3k d^3p ~ G(k)G(k+p)D(p) f^2(\k+\p/2) \times \nn \\
      &\qquad\qquad\qquad\qquad \times \left[\mathcal{K}(k+q,k)G(k+q)f^2(\k+\q/2) + \mathcal{K}(k-q,k)G(k-q)f^2(\k-\q/2)\right] \nn \\
      &= - \frac{N\gb}{\chi_0\tpp^6} \int d^3k d^3p ~ D(p) \times \left[ G(k)G(k+p)D(p) f^2(\k+\p/2) \mathcal{K}(k+q,k)G(k+q)f^2(\k+\q/2) + \right. \nn \\
      &\qquad\qquad\qquad\qquad\qquad\qquad\quad~~ \left. G(k+q)G(k+p+q)D(p) f^2(\k +\p/2 +\q)\mathcal{K}(k,k+q)G(k)f^2(\k+\q/2)\right] \nn \\
      &= - \frac{N\gb}{\chi_0\tpp^6} \int d^3k d^3p D(p) G(k)G(k+q)  \mathcal{K}(k+q)f^2(\k + \q/2) \times \nn \\
  &\qquad\qquad\qquad\qquad \times \left[G(k+p) f^2(\k+\p/2) - G(k+p+q)f^2(\k + \p/2 +\q) \right]
\end{flalign}

For $\q\to 0$ we get
\begin{align}
  \label{eq:I-d-2}
  I_d^{SE} = -\frac{N\gb}{\chi_0\tpp^6} \int d^3k d^3p ~ G(k)G(k+q)G(k+p)G(k+p+q)  \mathcal{K}(k+q;k) \mathcal{K}^{-1}(k+p+q;k+p) D(p) f^2(\k) f^2(\k + \p/2).
\end{align}
Summing up the two terms gives Eq. \eqref{eq:Id-def_a}, which does not assume anything about $\mathcal{K}$, i.e. is correct also for fermions with self-energy included.

\begin{figure}
  \centering
  \includegraphics[width=0.5\hsize,clip,trim=0 125 0 75]{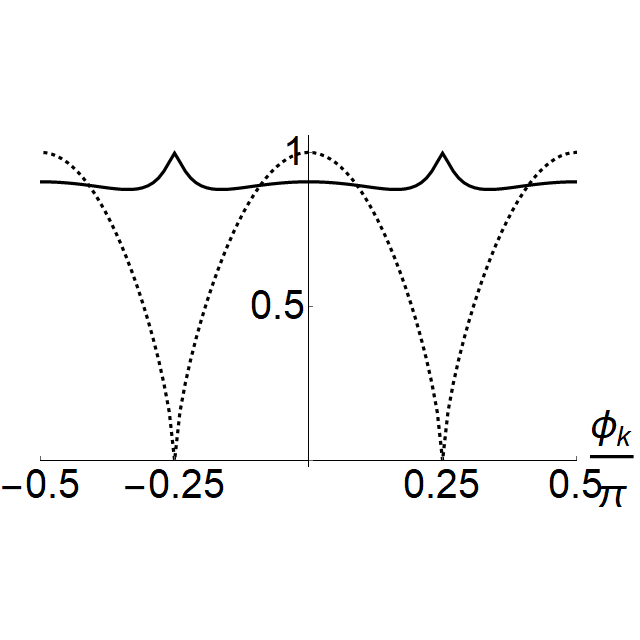}
  \caption{$f_k$ for a circular FS. The thick line is $f_k / (2\times 2^{2/3}\pi/3\sqrt 3)$, calculated numerically for $f(\phi_k) = \cos 2 \phi_k$. The dashed line is $|f(\phi_k)|^{2/3}$, included for comparison}
  \label{fig:supp-mu1-fig}
\end{figure}

When using the dispersion for free fermions we have,
\begin{equation}
  \label{eq:KK-free}
  \mathcal{K}(k+q;k) \mathcal{K}^{-1}(k+p+q,k+p) = [i q_0-\ve(\k+\q)+\ve(\k)][i q_0-\ve(\k+\p+\q)+\ve(\k+\p)]^{-1} = 1.
\end{equation}
The last equality is exact for any fermionic dispersion at $\q = 0$.
When using the bare theory, the $\mathcal{K}\mathcal{K}^{-1}$ term disappears and we can just expand the remaining angular function to second order in $q/\kf\sin\phi$, where $\cos\phi=\hat{k}\cdot\hat{q}$. Within Eliashberg theory, the expression factors into three parts, two fermionic and one bosonic. Each fermionic part is of the form
\begin{align}
  \label{eq:If-free}
  I_f = \int dk_0 d^2k G(k)G(k+q) &= \frac{\kf}{\vf}\int d k_0d\ve_kd\phi\frac{1}{i k_0 - \ve(\k)}\frac{1}{i (k_0+q_0) - \ve(\k)} \nn\\
  &\simeq (2\pi)^2i\kf/\vf \int dk_0 \frac{\Theta(k_0+q_0)-\Theta(k_0)}{i q_0}=(2\pi)^2\kf/\vf\int_{-q_0}^0\frac{dk_0}{q_0}.
\end{align}
Note that since $\q = 0$, the residue of the integration over the momentum transverse to the FS has no dependence on the momentum parallel to the FS. One of the frequency integrals can be done immediately. Then we are left with,
\begin{align}
  \label{eq:supp-Id-eliashberg-bare}
  I_d \propto \int_{-q_0}^0\frac{dp_0}{q_0}\int d\phi \frac{p(\phi) \mathcal{F}(\phi_k;\phi_k+\phi)}{p(\phi)^3+f^2\left(\theta(\phi_k,\phi)\right)\gamma |p_0|/\vf + c^{-2}p_0^2p(\phi)}
\end{align}
Here, as in Eq. \eqref{eq:lambda-def} and Eq. \eqref{eq:supp-sg0-explicit}, $p,\theta$ trace out the length and position of the bosonic momentum on the FS. $\mathcal F$ was defined in Eq. \eqref{eq:Fkk}. As usual, we split the integral into a static and dynamic part. In anticipation of the end result, we write down these parts as $\mu_0,\mu_1$ from Eq. \eqref{eq:mu-expr}, with the appropriate prefactors. The static part gives,
\begin{equation}
  \label{eq:supp-mu0-explicit}
  \mu_0f^2 = \int d\phi \frac{f(\phi_k)f^2(\phi_k+\phi/2)\left[f(\phi_k)-f(\phi_k+\phi)\right]}{p(\phi)^2}.
\end{equation}
For a circular FS we get
\begin{equation}
  \label{eq:mu0-circ}
  \mu_0f^2 \simeq 2.15\pi \cos^2(2\phi_k)(2\cos^24\phi_k-1), \quad\langle \mu_0 f^2 \rangle \simeq 1.69.
\end{equation}
Next, we add and subtract the static part from Eq. \eqref{eq:supp-Id-eliashberg-bare} to get the dynamic contribution,
\begin{align}
  \label{eq:supp-mu1-explicit}
  \frac{\gb}{\ef}\mu_1 = f_2(\phi_k) \int_{-\infty}^{\infty} d\phi \frac{f^2\left(\theta(\phi_k,\phi)\right)(\vf\kf^3)^{-1}\gamma |p_0| + (\kf c)^{-2}p_0^2\phi}{|\phi|^3 + f^2\left(\theta(\phi_k,\phi)\right)(\vf\kf^3)^{-1}\gamma |p_0|+ (\kf c)^{-2}p_0^2\phi},
\end{align}
Here, we used the convergence of the momentum integration to expand $\mathcal F$ to second order and approximate $p(\phi) \simeq \kf |\phi|$. Eq. \eqref{eq:supp-mu1-explicit} can be rescaled to give,
\begin{align}
  \label{eq:supp-mu1-explicit-2}
  \frac{\gb}{\ef}\mu_1 &= f_2(\phi_k) \left|\frac{\gamma p_0}{\vf\kf^3}\right|^{1/3}\int_{-\infty}^{\infty} d\phi \frac{f^2(\theta)+|p_0/\w_1|^{4/3}\phi}{|\phi|^3 + f^2(\theta)+|p_0/\w_1|^{4/3}\phi} \nn \\
  &=  \left|\frac{\gamma q_0}{\vf\kf^3}\right|^{1/3} f_2(\phi_k)f_k(|p_0|/\w_1)\left|\frac{p_0}{q_0}\right|^{1/3}.
\end{align}
Here, $\w_1^2 = \gamma c^3/\vf$ and $f_k$ is an interpolating function with the following limits,
\begin{equation}
  \label{eq:supp-fk-def}
  f_k(x) =  \left\{
    \begin{array}{cc}
      \int d\phi \frac{f^2(\theta)}{|\phi|^3 + f^2(\theta)}\simeq \frac{2\times 2^{2/3}\pi}{3\sqrt{3}} & |x|\ll 1 \\
      \frac{\pi }{2}|x|^{2/3} & |x| \gg 1
    \end{array}
\right. .
\end{equation}
See Fig. \ref{fig:supp-mu1-fig} for a depiction of $f_k$ in the low frequency limit $p_0 \ll \w_1$.
The $\mu_0,\mu_1$ given in Eqs. \eqref{eq:supp-mu0-explicit}+\eqref{eq:supp-mu1-explicit} are the $\mu_0,\mu_1$ from Eq. \eqref{eq:mu-def}, and these in turn give the constants $B, C$ in Eqs. \eqref{eq:pi-bare-wrong}+~\eqref{eq:pi-bare-wrong}, see Eq. \eqref{eq:pi-full-result_1}.

\subsection{Perturbative evaluation of $\Pi^{(1)}$ for $q_0 \ll \w_0$ with self-energy insertion and $f = 1$}
\label{sec:pert-eval-pi1}

Supposing we introduce $\Sg$ into the Green's functions, so now $\mathcal{K}\mathcal{K}^{-1}$ is no longer unity. Consider the case $f = 1$. Evaluating again in the Eliashberg approximation we find
\begin{align}
  \label{eq:supp-Id-nongauge}
  I_d \propto \int_{-q_0}^0dk_0\mathcal{K}^{-1}(k_0+q_0,k_0)\int_{-q_0}^0dk_0'\mathcal{K}^{-1}(k_0'+q_0,k_0')\frac{\mathcal{K}(k_0'+q_0,k_0')-\mathcal{K}(k_0+q_0,k_0)}{\mathcal{K}(k_0+q_0,k_0)}\times |k_0-k_0'|^{-1/3}.
\end{align}
Here, the $\mathcal{K}^{-1}$ factors have replaced the $q_0^{-1}$ factors in Eq. \eqref{eq:If-free}. Since the indices are arbitrary, we can symmetrize this expression and obtain
\begin{align}
  \label{eq:supp-Id-nongauge-2}
  I_d \propto \int_0^{q_0}dk_0\int_0^{q_0}dk_0'\frac{\left(\mathcal{K}'-\mathcal{K}\right)^2}{(\mathcal{K}\mathcal{K}')^2}\times |k_0-k_0'|^{-1/3}.
\end{align}
Thus, $I_d$ is by necessity non-zero. For $\Sg\sim \w^{2/3}$ we find that $I_d$ scales as $q_0^{1/3}$.

\section{Derivation of bosonic and fermionic self-energies within the Eliashberg framework}
\label{sec:derivation-eqs-pi-sig}

In this section we detail how to obtain the angular dependent behavior for $\Pi(q)$ and $\Sg(k)$, in the region $q_0 \ll \vf |\q|,k_0\ll\vf |\k|$, Eqs. \eqref{eq:pi-1-loop}+\eqref{eq:se=1-loop}. Similar expressions, except for the functions $\lambda_0,\lambda_1$ in Eq. \eqref{eq:lambda-def}, have been found before (see e.g. \cite{Oganesyan2001}). The expression for $\Pi$ is given in Eq. \eqref{eq:pi-def}. Shifting momentum and integrating over momentum transverse to the FS we find,
\begin{equation}
  \label{eq:pi-momentum}
  \Pi(q) = N\frac{\kf}{\tpp^2\vf}\int_{-q_0}^{0} dk_0 \int d\phi \frac{i~f^2(\phi_k)}{i\Sgt(k+q/2)-i\Sgt(k-q/2)-\vf q \cos(\phi-\phi_k)} \simeq N\frac{\kf}{\tp\vf}\frac{\q_0}{\vf |\q|}f^2(\phi_k + \pi/2)
\end{equation}

Next, we compute Eq. \eqref{eq:se=1-loop}. Starting from the definition of Eq. \eqref{eq:sigma-def}, we have
\begin{equation}
  \label{eq:supp-Sg-expr}
  \Sg(k) = \frac{\gb}{\tpp^3}\int dp_0 \int d^2p \frac{f^2(\k+\p/2)}{i\Sgt(k+p)-\ve(k+p)} \frac{|\p|}{|\p|^3+ f^2(\hat p')\gamma |p_0|/\vf}
\end{equation}
where $\hat p' = \hat z \times \hat p$. It is convenient to split $\Sg$ into a part that does not have $f$ terms varying in the integrand and a part where the variation of $f$ is taken explicitly into account. Adding and subtracting this part we find
\begin{equation}
  \label{eq:supp-sg-split}
  \Sg = \Sg_0 + \delta\Sg,
\end{equation}
where
\begin{align}
  \label{eq:supp-sg0}
  \Sg_0(k) &= \frac{\gb}{\tpp^3}\int dp_0 \int d^2p \frac{f^2(\hat k)}{i\Sgt(k+p)-\ve(k+p)} \frac{|\p|}{|\p|^3+ f^2(\hat k)\gamma |p_0|/\vf} \nn\\
           &\simeq \frac{\gb}{\tpp^3}\int dp_0 \int dp_\perp dp_\parallel\frac{f^2(\hat k)}{i\Sgt(k+p)-\vf (p_\perp+k_\perp)} \frac{\sqrt{p_\perp^2+p_\parallel^2}}{(p_\perp^2+p_\parallel^2)^{3/2}+ f^2(\hat k) \gamma |p_0|/\vf} \nn\\
           &= \frac{\gb}{\tpp^3}\int dp_0 \int dp_\perp \frac{f^2(\hat k)}{i\Sgt(k+p)-\vf (p_\perp+k_\perp)}\int dp_\parallel \frac{|p_\parallel|}{|p_\parallel|^3+ f^2(\hat k) \gamma |p_0|/\vf}\left(1 + O(k_0^{2/3})\right)\nn\\
           &=i\w_0^{1/3}|f(\hat k)|^{4/3}k_0^{2/3} + O(k_0^{4/3}).
\end{align}
The line before the last is the essential step of the Eliashberg approximation. In the regime where $k_\perp \sim p_\perp \sim \Sgt$, we may neglect $p_\perp$ in the bosonic propagator, up to order $p_\perp^2/p_\parallel^2 \sim k_0^{2/3}$, leading to the final line of Eq. \eqref{eq:supp-sg0}.

Next we compute $\delta \Sg$, which in explicit form is,
\begin{align}
  \label{eq:supp-dsg-1}
  \delta\Sg(k) &= \frac{\gb}{\tpp^3}\int dp_0 \int d^2p \frac{1}{i\Sgt(k+p)-\ve(k+p)} \times \Theta(\k,\p,p_0),
\end{align}
where
\begin{equation}
  \label{eq:supp-Theta-def}
  \Theta(\k,\p,p_0) = |\p| \left[\frac{f^2(\k+\p/2)}{|\p|^3+ f^2(\hat p')\gamma |p_0|/\vf} - \frac{f^2(\hat k)}{|\p|^3+ f^2(\hat k)\gamma |p_0|/\vf} \right].
\end{equation}

The frequency integration is dominated by the region $|\p|^3 \sim p_0 \ll \vf |\p|$, so we again evaluate it in the Eliashberg approximation, integrating out the fermionic sector first.. Again, it is convenient to split $\delta\Sg/k_0$ into a static and dynamic part. The static part is just,
\begin{align}
  \label{eq:supp-dsg-static}
  \delta\Sg(\hat k)_0 = \frac{\gb}{\tpp^3}\int dp_0 \int d^2p \frac{1}{i\Sgt(k+p)-\ve(k+p)} \times |\p| \left[\frac{f^2(\k+\p/2)-f^2(\hat k)}{|\p|^3} \right].
\end{align}
Evaluating the fermionic part first we get,
\begin{align}
  \label{eq:supp-sg0-explicit}
  i\delta\Sg(k)_0 = \frac{\gb N_F}{\tpp^2}k_0\int d\phi \frac{f^2(\phi_k+\phi/2)-f^2(\phi_k)}{p(\phi)^2}.
\end{align}
Here, $p(\phi)$ traces out the length of the bosonic momentum, stretching from the FS at $\phi_k$ to the FS at $\phi_k+\phi$, where $\phi$ goes around the unit circle. Eq. \eqref{eq:supp-dsg-static} yields the $\lambda_0$ term of Eq. \eqref{eq:lambda-def}. Thus, e.g. for a circular FS, $p(\phi) = 2k_f|\sin(\phi/2)|$, and the result of the integral is
\begin{equation}
  \label{eq:supp-lambda0-circle}
  \lambda_0 \simeq -2.15\pi\cos(4\phi_k)
\end{equation}
Finally we compute the part of $\delta\Sg$ that depends on $k_0$ in a nonlinear way. After some manipulations of Eq. \eqref{eq:supp-dsg-1} we find,
\begin{align}
  \label{eq:supp-dsg-2}
  \delta\Sg(k)_2 &= \frac{\gb}{\tpp^3}\int dp_0 \int d^2p \frac{1}{i\Sgt(k+p)-\ve(k+p)} \times  \nn\\
  &\qquad\qquad \times |\p|
  \left[\frac{\left(f^2(\k+\p/2)-f^2(\hat k)\right)
    \left[1 - \left(1+\frac{f^2(\hat p)\gamma |p_0|}{|\p|^3}\right)\left(1+\frac{f^2(\hat k)\gamma |p_0|}{|\p|^3}\right)\right]}{|\p|^3\left(1+\frac{f^2(\hat k)\gamma |p_0|}{|\p|^3}\right)\left(1+\frac{f^2(\hat p)\gamma |p_0|}{|\p|^3}\right)} + \right.\nn\\
  &\qquad\qquad\qquad\qquad \left.
    \frac{f^2(\hat k)\gamma|p_0|\left(f^2(\hat p)-f^2(\k+\p/2)\right)}{|\p|^6\left(1+\frac{f^2(\hat k)\gamma |p_0|}{|\p|^3}\right)\left(1+\frac{f^2(\hat p)\gamma |p_0|}{|\p|^3}\right)}\right]
\end{align}
Eq. \eqref{eq:supp-dsg-2} yields $\lambda_1$ in Eq. \eqref{eq:lambda-def}. The second term in the bracket has a form factor part that is $O(1)$ even for $|\p| \sim 0$. As a result it contributes even in the regime $|\p| \sim p_0^{2/3}$, which is formally not within the Eliashberg approximation that assumes $|\p| \sim p_0^{1/3}$. However the only effect of this is some modification of the functional form of $\lambda_1$, which is not overly important for this work.

\section{The Eliashberg density vertex and the results of sec. \ref{sec:calculating-gamma-f}}
\label{sec:eliashb-vert-gamm}

In this section we evaluate the density vertex function for an incoming boson of small momentum within Eliashberg theory:
\begin{equation}
  \label{eq:supp-gamma-1}
  \Gamma_0(k;q) = \Gamma_0(k_0, \k; q_0, \q).
\end{equation}
Here as usual $k;q$ respectively denote the fermionic and bosonic degrees of freedom. We used these results in secs. \ref{sec:breakd-eliashb-theor} and \ref{sec:first-order-corr}. Our derivation generalizes the approach of Ref. \onlinecite{Chubukov2005}.

In calculations done within the Eliashberg regime, one may generally take $|\k|-\kf$, to be zero. This is because internal fermionic degrees of freedom are within a distance $q_f \sim\w^{2/3}/N^{1/3}$ of the FS, whereas bosonic internal degrees of freedom are at at a larger distance $q_b \sim (N\w)^{1/3}$. By dimensional analysis we expect the vertex to depend on $q_b,q_f/q_b \sim 0$. When we go beyond the Eliashberg regime we expect $q_f/q_b \sim 1$ and should be more careful. In this section we will provide general expressions for the vertex, and specialize them to the relevant regions we used in the manuscript.

\begin{figure}
  \centering
  \includegraphics[width=0.4\textwidth,clip,trim=100 500 100 100]{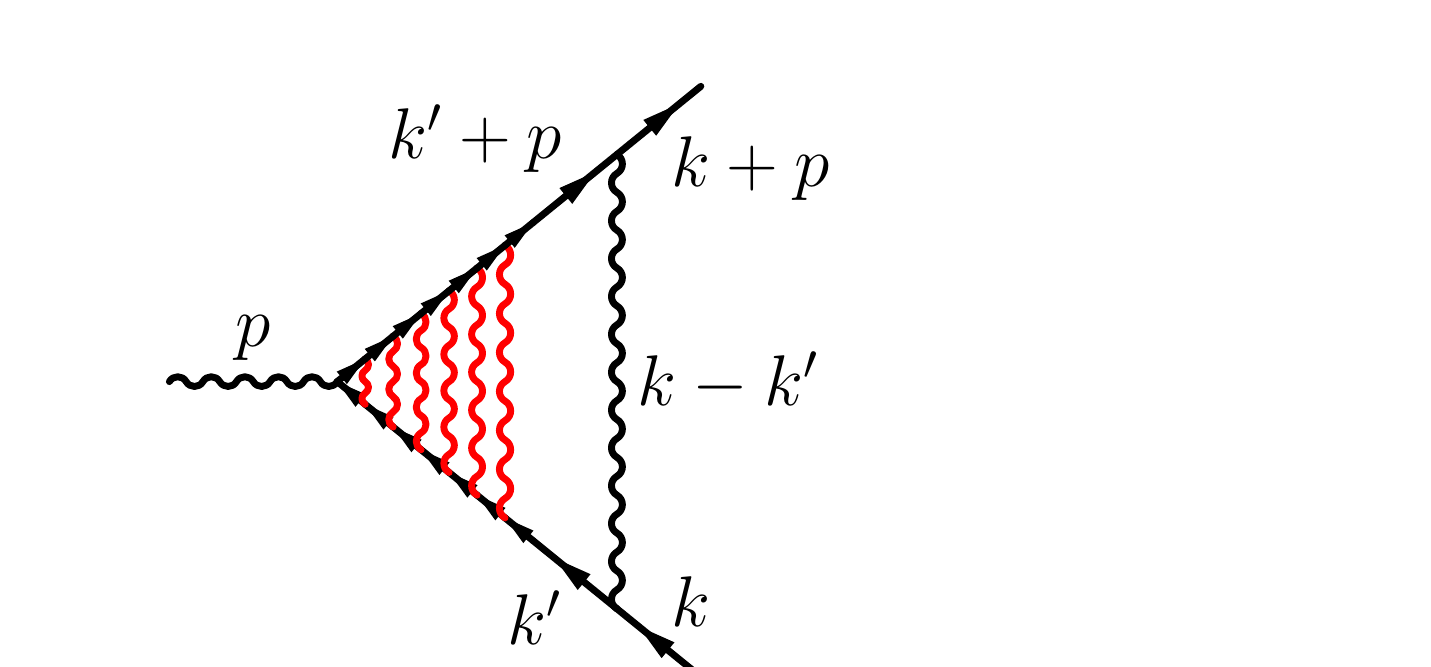}
    \includegraphics[width=0.4\textwidth,clip,trim=100 500 100 100]{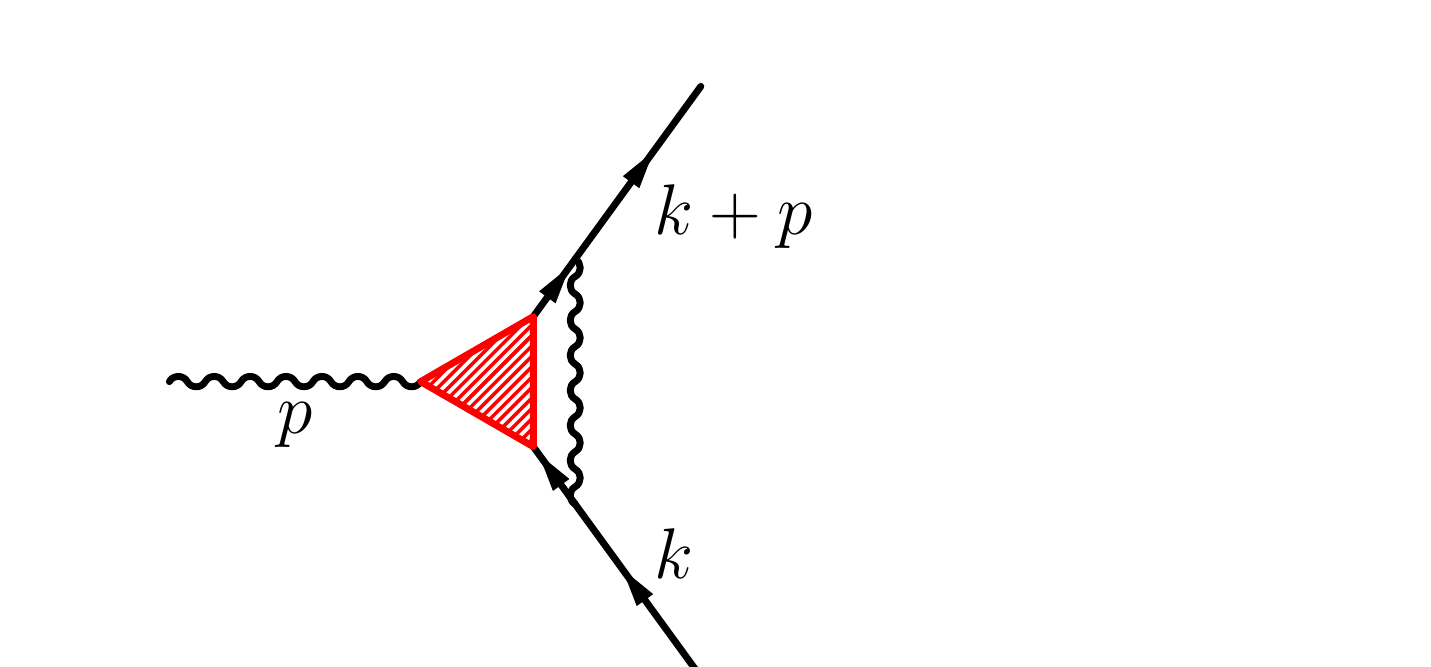}
  \caption{Ladder form of the vertex. (left panel) a sample diagram in the ladder expansion for the vertex $\Gamma_0(k_0,\k; p_0, \p)$. The momentum $\k$ only appears on the outside legs and does not penetrate into the internal ladder rungs. This as opposed to $\p$ which shows up on every rung. The red bosonic lines depicts internal lines belonging to segments that can be evaluated with momenta on the $FS$ itself. Only the last rung of the ladder is forced, by the external legs, to have weight at a distance $q_\perp - \kf$ from the $FS$. (right panel) the effective vertex correction. The red vertex is evaluated for $\k$ on the FS itself.}
  \label{fig:vertex-ladder}
\end{figure}

From Fig. \ref{fig:coupling-vertex}, the integral equation for the full vertex $\Gamma$ is given by Eqs. \eqref{eq:Gamma-dyson}+\eqref{eq:delta-Gamma-def}. Explicitly it is:
\begin{align}
  \label{eq:gamma-int-def}
  \delta\Gamma(k;q) &= \frac{\gb}{\tpp^3}\int d^3p \frac{\Gamma(p; q)}{(i\Sgt(p_0+q_0) - \ve(p)-\vf\hat{p}\cdot \q)(i\Sgt(p+q) - \ve(p)-\vf\hat{p}\cdot \q)}\frac{ f(\hat{k})^{-1} f^2\left(\frac{\k+\p}{2}\right)|\p-\k|}{|\p-\k|^3+\gamma f^2(\hat{p}')|p_0-k_0|/\vf}
\end{align}
From this point on we choose choose $\q = q\hat{x}$, and assume implicitly that $\hat k \simeq \hat x$. This choice allows us to avoid issues related with FS curvature when $\hat k \cdot \q \sim q_0^{2/3}$. Next, we specialize to the density vertex $\Gamma_0$. This is equivalent to dropping the internal angular dependence of $\Gamma$ with the integral. We integrate over $d\ve_p =dp_x \kf/\vf $ in the fermionic sector and find
\begin{align}
  \label{eq:supp-gamma-3}
  \delta\Gamma_0(k;q) &= \frac{\gb}{\tpp^2}\int_{-q_0}^0 dp_0\frac{\Gamma_0(p_0,0;q_0,q)}{\Sgt(p+q)-\Sgt(p)+i\vf q}\int dp \frac{(p^2+k_x^2)^{1/2}}{(p^2+k_x^2)^{3/2}+\gamma f^2(\hat{k})|p_0-k_0|/\vf} \nn\\
  &=\frac{\gb}{\tpp^2}\int_{-q_0}^0 dp_0\frac{\Gamma_0(p_0,0;q_0,q)}{\Sgt(p+q)-\Sgt(p)+i\vf q} \times \left[\left(\frac{\vf f^4(\hat{k})}{\gamma|p_0-k_0|}\right)^{1/3}\int_0^{\infty} d\phi \frac{(\phi^2+\kappa^2)^{1/2}}{(\phi^2+\kappa^2)^{3/2}+1} + \frac{d\tilde\mu}{dp_0}\right]
\end{align}
where we defined $\kappa^3 = \vf k_x^3/\gamma f^2(\hat{k})|p_0-k_0|$, and
\begin{align}
  \label{eq:mu-exact-def}
  \frac{d\tilde\mu}{dp_0} = \kf\int d\phi~ \Theta(\k,\p,p_0-k_0).
\end{align}
Here, $\phi$ is an integration along the FS, and $\Theta$ was defined in Eq. \eqref{eq:supp-dsg-1}. The $d\phi$ integral in Eq. \eqref{eq:supp-gamma-3} defines a function with the following asymptotics,
\begin{equation}
  \label{eq:f-bosonic-1}
  g(y) = \int_0^{\infty} dx \frac{(x^2+y^2)^{1/2}}{(x^2+y^2)^{3/2}+1} = \left\{
  \begin{array}{cc}
    \frac{\pi}{2y} & y \gg 1 \\
    \frac{2\pi}{3\sqrt{3}} & y \ll 1
  \end{array}
\right.
\end{equation}
Within our treatment, and neglecting $\mu$, the vertex correction depends only on the momentum transfer perpendicular to the FS and on frequency. Note that in Eq. \eqref{eq:supp-gamma-3} the internal $\Gamma_0$ no longer depends on $\q$ \emph{or} $\k$. The reason for this can be seen by looking at Fig. \ref{fig:vertex-ladder}. Just by placing internal integration variables on the ``rungs'' of the ladder, it is easy to see that even for $\k \gg k_0^{2/3},p_0^{2/3}$ the internal rungs still contribute from regions close to the FS. Only the last rung depends strongly on the external legs. Physically, the interpretation is that even for excitations perpendicular to the FS, it is possible to excite a large cloud of virtual particle-hole pairs by making a single virtual transfer to the vicinity of the FS.
In any case, we see that for any $0 < \kappa < 1$ the bosonic sector scales as $[ \gamma f^2(\hat{k})|q_0-k_0|/\vf]^{-1/3}$, up to corrections of order 1. So, for all calculations performed henceforth (and used in the body of the manuscript) we take $g(\kappa) = g(0)$. Combining factors together we end up with
\begin{align}
  \label{eq:supp-delta-gamma-freq}
  \delta \Gamma_0 &= \frac{2}{3}\int_{-q_0}^{0}dp_0 \frac{\Gamma_0(p_0,0;q_0,q)}{\Sgt(p+q)-\Sgt(p)+i\vf q} \times \left|\frac{\w_0f^4(\hat{k})}{p_0-k_0}\right|^{1/3} \nn\\
  &\qquad+ \frac{\gb f(\hat{k})}{\tpp^2}\int_{-q_0}^0 dp_0\frac{\Gamma_0(p_0,0;q_0,q)}{\Sgt(p+q)-\Sgt(p)+i\vf q} \frac{d\tilde\mu}{dp_0}
\end{align}
This is just a more generic form of Eq. \eqref{eq:dg-eliashberg-form}. Eq. \eqref{eq:supp-delta-gamma-freq} can be solved generically via the ansatz,
\begin{equation}
  \label{eq:supp-gamma-ansats}
  (iq_0 - \vf \hat{k}\cdot \q)\Gamma_0(k;q) = \left[i\Sgt(k+q)-i\Sgt(k)] - \vf\hat{k}\cdot q\right] = G^{-1}(k+q)-G^{-1}(k).
\end{equation}
which is just a Ward identity. However, recall that that Eq. \eqref{eq:supp-gamma-ansats} assumes implicitly that $\hat{k}\cdot\hat{q} \simeq 1$.
Another implication of the Ward identity, is that far away from the FS, the fermionic $\w^{2/3}$ self energy is cut off by the same scale that cuts of the vertex correction, i.e.,
\begin{equation}
  \label{eq:app-SE-cutoff}
  \Sg(k, k_0) \sim \Sg(k_0) g\left(\frac{k}{(\gamma f^2(\hat k) |k_0|/\vf)^{1/3}}\right),
\end{equation}
so it vanishes as $k^{-1}$ in the large $k$ limit.

The leading ordere form of the full vertex was already derived in the manuscript itself, except for an explicit form of $\mu_0,\mu_1$. Expressions for these appear in the previous section, Eqs. \eqref{eq:supp-mu0-explicit}+\eqref{eq:supp-mu1-explicit}.

\section{Derivation of Eq. \eqref{eq:Id-correct-1} in the rung-by-rung analysis of sec. \ref{sec:deriving-rung-rung}}
\label{sec:deriv-results-sec}

In this section we derive Eq. \eqref{eq:Id-correct-1}, the expression for the polarization using fully renormalized side density vertices and self-energies. To do so we evaluate the diagrams of Fig. \ref{fig:full-three-diags}, but now with fully dressed vertices and Green's functions. The expressions for the diagrams are,

\begin{align}
  \label{eq:supp-I-pm-dressed}
  I_\pm &= \frac{N\gb}{\chi_0\tpp^6}\int d^3k d^3p ~ G^2(k)G(k+p)G(k \pm q) D(p) \Gamma_0^2(k;\pm q)f^2(\k)f^2(\k+\p/2) \\
  \label{eq:supp-I-v-dressed}
  I_v &= \frac{N\gb}{\chi_0\tpp^6}\int d^3k d^3p ~ G(k)G(k + q)G(k+p)G(k+p+q)D(p) \times \nn \\
  &\qquad\qquad\qquad\qquad \times \Gamma_0(k;q)f(\k)\Gamma_0(k+p;q)f(\k+\p)f^2(\k+\p/2)
\end{align}
Here, we used the identity
\begin{equation}
  \label{eq:supp-Gamma-relation}
  \Gamma_0(k-q;q) = \Gamma_0(k;-q)
\end{equation}
which is a result of the Ward identity Eq. \eqref{eq:supp-gamma-ansats}, and the fact that vertex corrections on the internal vertices are small. We also use full Green's functions, with the self-energy taken as the first term in Eq. \eqref{eq:se=1-loop}. The identity in Eq. \eqref{eq:GG-kappa} can now be written in the form,
\begin{equation}
  \label{eq:supp-Gamma-Green-rel}
  \Gamma_0(k;q)G(k+q)G(k) = \frac{\Gamma_0(k;q)}{\Sgt(k+q)-\Sgt(k)}\left[G(k) - G(k+q)\right] = \frac{1 }{i q_0-\vf \hat{k}\cdot \q}\left[G(k) - G(k+q)\right]
\end{equation}
The static part of the polarization is computed by taking $q_0 \to 0$. In this limit vertex corrections are negligible, and thus the static part is still described by Eq. \eqref{eq:app-I-s-1}. The dynamical part of the self energy for $\q = 0$ is now
\begin{align}
  I_d^{SE} &= \frac{N\gb}{\chi_0\tpp^6}\int d^3k d^3p ~ G(k)G(k+p) D(p) f^2(\k+\p/2) \frac{\Gamma_0(k-q;q)f(\k)G(k-q)-\Gamma_0(k;q)f(\k)G(k+q)}{i q_0} \nn\\
  &= \frac{N\gb}{\chi_0\tpp^6}\int d^3k d^3p ~ G(k)G(k+q) D(p) f^2(\k+\p/2) \Gamma_0(k;q)(-1)\Gamma_0(k_0+p_0,\k+\p;q_0,\q=0)G(k+p+q)G(k+p)
\end{align}
Here, in the first line, we shifted the $k$ integral of the first term $k \to k+q$, and then used Eq. \eqref{eq:supp-Gamma-Green-rel}. By symmetrizing via $dp\to dk'=d(k+p)$ we end up with Eq. \eqref{eq:Id-correct-1}. Then, as in sec. \ref{sec:pert-eval-pi}, the integral splits into three: an integral over $k_\perp$, and integral over $k_\perp'$ and an integral over $k_\parallel,k_\parallel'$. Each transverse integral limits the frequency regime to $(-q_0,0)$,
\begin{equation}
  \label{eq:supp-gamma-gg-int}
  \int d^3k \Gamma_0(k;q) G(k) G(k+q) = \int d^3k \frac{G(k+q)-G(k)}{i q_0} \simeq \pi \vf^{-1}\int_{-q_0}^0\frac{dk_0}{q_0}\int dk_\parallel
\end{equation}
One of the frequency integrals can be performed immediately. The remaining integrals yield Eq. \eqref{eq:Pi-full-2nd}.

One last task is to check whether the dynamic part of $\Pi$ is ever negative, i.e. whether $\langle f_2 \rangle$ is ever positive. To check this, note that we can define
\begin{equation}
  \label{eq:fn-def}
  g_n(x) = f^2(\phi)\left[f'^2(\phi)+\frac{1}{n-1}f(\phi)f''(\phi)\right] = \frac{1}{n(n-1)}f^{4-n}(\phi)\frac{d^2}{d\phi^2}f^n(\phi),
\end{equation}
for $n=2,3,4$. We then have
\begin{equation}
  \label{eq:f-g}
  g_3(x) = f_2(x).
\end{equation}
We will use $g_2,g_4$ which are simple to evaluate, so as to get an expression for $\langle g_3 \rangle$. It is clear that
\begin{equation}
  \label{eq:g-4-avg}
  \langle g_4 \rangle = 0,
\end{equation}
since $g_4$ is a full derivative. In addition,
\begin{equation}
  \label{eq:g-2-avg}
  \langle g_2 \rangle < 0.
\end{equation}
To see this, note that $f(\phi)$ is periodic, and hence so is $f^2(\phi)$, so we may expand it in a Fourier series,
\begin{equation}
  \label{eq:f-sqr-fourier}
  f^2(\phi) = \sum_n (f^2)_n e^{2\pi i n \phi}.
\end{equation}
Then averaging $g_2$ yields
\begin{equation}
  \label{eq:g-2-avg-fourier}
  \int \frac{d\phi}{\tp} \frac{1}{2}f^2 (f^2)'' = \sum_{n,m} \frac{1}{2}(f^2)_n [-(2\pi m)^2] (f^2)_m \delta_{m,-n}= - 2\pi^2\sum_n n^2 |(f^2)_n|^2 < 0.
\end{equation}
Finally,
\begin{equation}
  \label{eq:g234}
  \langle f^3(\phi)f''(\phi)\rangle = \frac{3}{2}\langle g_2(\phi) - g_4(\phi) \rangle = \frac{3}{2}\langle g_2 \rangle
\end{equation}
Here we subtracted the expressions in Eq. \eqref{eq:fn-def} from one another. Adding all this together we find
\begin{equation}
  \label{eq:f2-avg}
  \langle f_2(\phi) \rangle = \langle g_3 \rangle = \langle g_2 - \frac{1}{2}f^3 f''\rangle = \langle g_2 - \frac{3}{4}g_2 \rangle = \frac{1}{4}\langle g_2 \rangle < 0.
\end{equation}
Therefore, $\langle f_2 \rangle < 0$ is always negative, so the nonconstant part of the polarization is always positive for $\q = 0, q_0 \neq 0$.

\section{Aslamazov-Larkin  (AL) diagrams for $\Pi (\q=0,q_0)$}
\label{sec:aslam-lark-corr}

\begin{figure}
  \centering
  \subfloat{
    \includegraphics[width=0.33\textwidth,clip,trim=120 500 150 120]{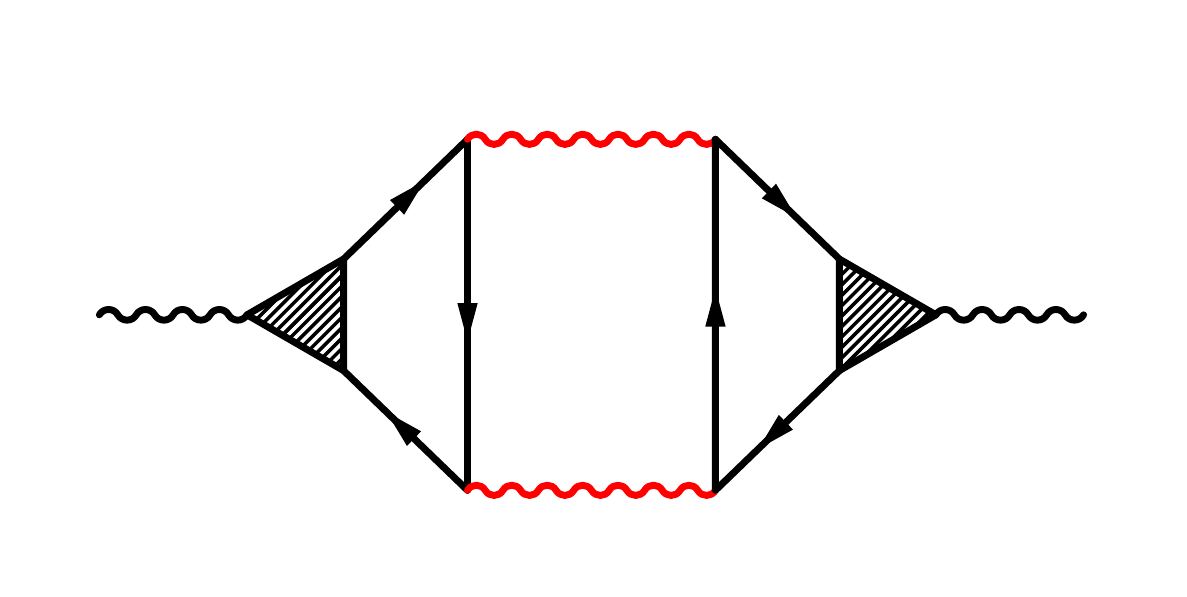}}
  \subfloat{
    \includegraphics[width=0.33\textwidth,clip,trim=120 500 150 120]{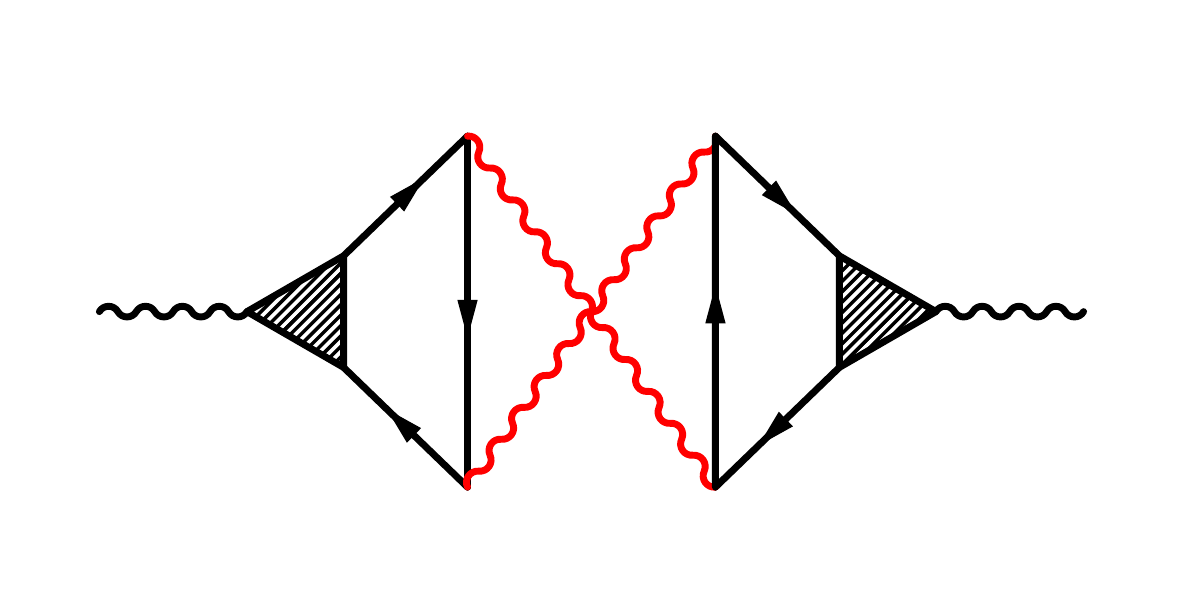}}

  \caption{Higher order Aslamazov-Larkin diagrams contributing to the dynamic polarization. For coupling in the charge channel these cancel out, and for coupling in the spin channel they contribute in the same order as the vertex and self-energy diagrams of Fig. \ref{fig:full-three-diags}. The red lines are the ones where form-factor variation is taken into account.}
  \label{fig:AL-diagrams}
\end{figure}
In the body of the manuscript, we pointed out that the contribution from the Aslamazov-Larkin diagrams, shown in Fig. \ref{fig:3-loop} in the regime $q_0 \ll \w_1$, are of the same order in $\gb$ and $N$ as the self-energy and vertex corrections of Fig. \ref{fig:2-loop} (see Ref. \onlinecite{Chubukov2014}). In this appendix we review how this result comes about, evaluate the diagrams, and show that they only serve to modify $\mu_0,\mu_1$ that we found in Eqs. \eqref{eq:supp-mu0-explicit} + \eqref{eq:supp-mu1-explicit}.

We treat the AL contributions within the Eliashberg theory for $q_0 \ll \w_0$, as all results for frequencies above $\w_0$ follow directly from the Eliashberg treatment. In the same manner as for the two-loop diagrams, the side vertices of the AL diagrams are dressed with density vertices $\Gamma_0$, as depicted in Fig. \ref{fig:AL-diagrams}.

The contribution from the AL diagrams is:
\begin{align}
  \label{eq:AL-def}
  \Pi_{AL}(q) &= -\frac{N^2 \gb^2}{\chi_0^2\tpp^9}\int d^3k d^3p d^3k' f(\k)f^2(\k+\p/2)\Gamma_0(k;q)G(k)G(k+q)G(k+p)D(p-q)D(p) \times\nn\\
  &\qquad \times \left[G(k')G(k'+q)G(k'+p)f^2(\k'+\p/2)+G(k'+q)G(k')G(k'+q-p)f^2(\k'-\p/2)\right]\Gamma_0(k';q)f(\k').
\end{align}
Here, we shifted $k'\to k'+q$ in the right-hand AL diagram. The key to simplify Eq. \eqref{eq:AL-def} is to note that $\chi_0D^{-1}(p) = |\p|^2 +\gb\Pi(p)$, so that the two bosonic propagators can actually be factorized:
\begin{equation}
  \label{eq:chi-factor}
  D(p)D(p-q) = \chi_0[\gb\Pi(p)-\gb\Pi(p-q)]^{-1}[D(p-q) - D(p)].
\end{equation}
Next, we use the Ward identity, Eq. \eqref{eq:gamma-0-sol}, to obtain,
\begin{align}
  \label{eq:supp-ggg-pp}
  \int \frac{d^3k}{\tpp^3} \Gamma_0(k;q)G(k)G(k+q)G(k+p)f(\k)f^2(\k+\p/2) &= \frac {1}{\tpp^3i q_0}\int  [G(k)-G(k+q)]G(k+p) f(\k)f^2(\k+\p/2).
\end{align}
In what follows, we should properly symmetrize Eq. \eqref{eq:AL-def} by inserting appropriate shifts of $p/2,q/2$. This is because it is necessary to track various angular expressions more accurately than for the 2-loop diagrams. To conserve space we do not do so explicitly, and just point out where it is necessary. Now, $\p$ is an internal integration variable so we can neglect vertex corrections involving $\p$. Therefore,
\begin{align}
  \label{eq:supp-gg-p-f}
  &N\gpref\int d^3k G(k)G(k+p)f(\k)f^2(\k+\p/2)\nn\\
  &\qquad\to N\frac{\gb\kf}{\tpp^2\vf} \int dk_0 d\phi f(\kf\hat{k}-\p/2) f^2(\kf\hat{k}) \frac{i\Theta(k_0'+p_0/2)-i\Theta(k_0'-p_0/2)}{ i\Sgt(k+p/2) - i\Sgt(k-p/2) -\vf |\p| \cos(\phi_p-\phi_k)} \nn \\
  &\qquad\simeq \gamma \frac{|p_0|}{\vf|\p|}f^2(\hat p') [f(\kf\hat{z}\times\hat{p}-\p/2) + f(\kf\hat{p}\times\hat{z}-\p/2]/2 \nn\\
  &\qquad= \gb\Pi(p) \tilde f(\p).
\end{align}
where in the second line we performed the symmetrization shift. The two terms in $\tilde f(\p)$ arise from the fact that the angular integration in Eq. \eqref{eq:supp-gg-p-f} has two peaks on opposite sides of the FS. Thus, the integration over the $d^3k$ variables gives a factor of $\tilde f(\p)[\gb\Pi(p)-\gb\Pi(p-q)]$. Then we are left with,
\begin{align}
  \label{eq:supp-AL-1}
  \Pi_{AL}(q) &= -N\frac{\gb}{\chi_0\tpp^6}\int~d^3k' d^3p \tilde{f}(\p)\left[D(p-q)-D(p)\right]\times\nn\\ &\qquad\qquad G(k')G(k'+q)\left[G(k'+p)f^2(\k'+\p/2)+G(k'+q-p)f^2(\k'-\p/2)\right]\Gamma_0(k';q)f(\k'). \nn\\
              &= -N\frac{\gb}{\chi_0\tpp^6}\int~d^3k' d^3p \tilde{f}(\p) D(p) G(k')G(k'+q) \Gamma_0(k';q)f(\k')\times \nn\\
  &\qquad\qquad \left[\left(G(k'+p+q)-G(k'+p)\right)f^2(\k'+\p/2) + \left(G(k'-p)-G(k'-p+q)\right)f^2(\k'-\p/2)\right]
\end{align}
Eq. \eqref{eq:supp-AL-1} shows that indeed the AL contribution is of the same order as the 2-loop contributions in both $\gb$ and $N$. Finally we obtain:
\begin{align}
  \label{eq:supp-AL-2}
  \Pi_{AL}(q) &= -N\frac{\gb}{\chi_0\tpp^6}\int~d^3k' d^3p D(p) G(k')G(k'+q)\Gamma_0(k';q)\frac{G(k'+p+q)-G(k'+p)}{i q_0}f(\k')f^2(\k'+\p/2) \delta\tilde f(\p) \nn\\
  &= + N\frac{\gb}{\chi_0\tpp^6}\int~d^3k' d^3p D(p) G(k')G(k'+q)\Gamma_0(k';q)G(k'+p+q)G(k'+p)\mathcal F'(\k';\k'+\p)
\end{align}
where
\begin{equation}
  \label{eq:Fkk-prime-def}
  \mathcal F'(\k';\k'+\p) = f(\k')f^2(\k'+\p/2) \delta\tilde f(\p),
\end{equation}
and
\begin{equation}
  \label{eq:delta-tilde-f}
  \delta \tilde f(\p) = \tilde f'(\k',\p) - \tilde f'(-\k',-\p)
\end{equation}
with
\begin{equation}
  \label{eq:supp-fp-prime}
  \tilde f'(\p) = \frac{1}{2}\left[f(\k'+\p/2) + f(-\k'-3\p/2)\right]
\end{equation}
To derive Eq. \eqref{eq:delta-tilde-f} used the relationship $\phi_p = \phi_k + \pi/2 + \phi/2$, and reversed the symmetrizing shift. We see that Eq.~\eqref{eq:supp-AL-2} is an analogue of Eq. \eqref{eq:Id-correct-1}, with a somewhat different angular component. For a purely even form factor, as for the nematic one, the contribution is zero within our approximations.

\section{Polarization operator $\Pi (\q-0, q_0)$ beyond Eliashberg theory}
\label{sec:first-order-corr}

In this final appendix we compute the leading contribution to $\Pi (\q=0, q_0)$ beyond Eliashberg theory. Within the Eliashberg treatment we factorized the momentum integration, namely we integrated transverse to the FS in fermionic propagators and neglected the transverse momentum component in the bosonic susceptibility, i.e., approximated $D (k)$ by its value between the points on the FS. This approximation definitely works for the leading, frequency-independent term in Eq.  (\ref{eq:pi-full-result}) because it comes from parallel momenta of order $k_F$ and transverse momenta of order $q^{2/3}_0$ (this is the only option to avoid $q_0$ to a positive power in the overall factor). However, it is not a'priori guaranteed that within this approximation one gets the leading frequency dependence of $\Pi (\q=0, q_0)$.

To verify whether the factorization of momentum integration is justified for the frequency-dependent part of $\Pi (\q=0, q_0)$, we again repeat the procedure used in Sec.  \ref{sec:deriving-rung-rung}. Namely, we select one segment from which we pick up the contribution with the gradient of the form factor. In all other segments we neglect the variation of the form-factor between incoming and outgoing momenta of the interaction terms.  However, as opposed to the Eliashberg treatment, we do not factorize the momentum integration in the segment with the gradient of $f (\k)$.

One may readily verify that beginning with Eq. \eqref{eq:Id-correct-1}, instead of Eq. \eqref{eq:Pi-full-2nd} we end up with\begin{align}
  \label{eq:Ik-1}
  \Pi (\q=0, q_0)&\simeq \frac{\gamma }{\tpp^3}\int\frac{d\theta}{\tp}f_2(\theta)\int_{q_0<|k_0|}\frac{dk_0}{q_0}\int d^2k  \frac{\Sgt(k_0+q_0)-\Sgt(k_0)}{(i\Sgt(k_0+q_0) - \vf k \cos\phi)(i\Sgt(k_0) - \vf k \cos\phi)}\times \nn\\
  &\hspace{300pt}\frac{k(k^2\sin^2\phi)/\kf^2}{k^3 + f^2(\theta+\phi)\gamma|k_0 +q_0|/\vf k^3_F }.
\end{align}
If we factorize the momentum integration into integration over $k_\perp = k \cos{\phi}$ in the fermionic propagators and over $k_\parallel = k \sin{\phi}$ in the bosonic propagator, we reproduce Eq.  (\ref{eq:pi-full-result}).  If, instead, we subtract from the r.h.s. of (\ref{eq:Ik-1}) the constant term and in the remaining part do not factorize but rather assume that $k_\parallel$ and $k_\perp$ are of the same order, i.e., that typical $\phi$ are of order one, we find that typical $k$ in the integrand are of order $k^{1/3}$,  typical $k_0$ are of order $q_0$, and the frequency dependence of $\Pi (\q=0, q_0)$ is in the form $q^{2/3}_0$.  In explicit form the $q^{2/3}_0$ term, which we label a ${\tilde \Pi}$, is
\begin{align}
  \label{eq:pi-final-prefactors}
  \gb\tilde{\Pi}(\q=0, q_0) &\sim  \frac{\gb\gamma\w_0}{\ef^2}\langle f_2 \rangle  \left|\frac{q_0}{\w_0}\right|^{2/3} \nn\\
  &\sim \left(\frac{q_0}{D}\right)^{1/3} \times \left(\Pi (\q=0, q_0) - \Pi (\q,0)\right)
\end{align}
where $D \sim \ef^2 N^2 /{\bar g} \gg \Lambda$.  Hence, within low-energy theory (energies are smaller than $\Lambda$), the frequency dependence coming from the integration range where internal momenta along and transverse to the FS are of the same order, is much weaker than the one coming from the range where momenta transverse to the FS are much smaller than the ones along the Fermi surface.

\end{document}